\documentclass[twocolumn]{aastex62}

\usepackage{graphicx}
\usepackage{natbib}
\usepackage{gensymb}

\hypersetup{linkcolor=blue,citecolor=blue,filecolor=blue,urlcolor=blue}

\usepackage{amsmath}
\usepackage{colortbl}

\newcommand{\HI}{H{\sc\ i}}

\newcommand{\SiII}{Si{\sc\ ii}}
\newcommand{\SiIII}{Si{\sc\ iii}}
\newcommand{\SiIV}{Si{\sc\ iv}}
\newcommand{\OVI}{O{\sc\ vi}}
\newcommand{\CIV}{C{\sc\ iv}}
\newcommand{\CII}{C{\sc\ ii}}

\newcommand{\NV}{N{\sc\ v}}
\newcommand{\CaII}{Ca{\sc\ ii}}
\newcommand{\NaI}{Na{\sc\ i}}

\newcommand{\NeVII}{Ne{\sc\ vii}}
\newcommand{\NeVIII}{Ne{\sc\ viii}}
\newcommand{\OVII}{O{\sc\ vii}}
\newcommand{\OVIII}{O{\sc\ viii}}

\newcommand{\msun}{\rm M_\odot}
\newcommand{\msunyr}{\rm M_\odot~yr^{-1}}
\newcommand{\kms}{{\rm km \, s^{-1}}}

\newcommand{\rvir}{\rm R_{200}}
\newcommand{\vlsr}{v_{\rm LSR}}
\newcommand{\vgsr}{v_{\rm GSR}}
\newcommand{\NHI}{N_{\rm HI}}
\newcommand{\NOVI}{N_{\rm OVI}}

\usepackage{verbatim}

\shorttitle{FOGGIE III: The Mocky Way}
\shortauthors{Zheng et~al.}

\begin{document}
\defcitealias{peeples19}{Paper~I}
\defcitealias{corlies18}{Paper~II}
\defcitealias{blandhawthorn16}{BG16}
\defcitealias{werk14}{Werk14}

\title{Figuring Out Gas \& Galaxies in Enzo (FOGGIE). III.\\ The Mocky Way: Investigating Biases in Observing the Milky Way's Circumgalactic Medium}

\correspondingauthor{Yong Zheng}
\author[0000-0003-4158-5116]{Yong Zheng}
\affiliation{Department of Astronomy, University of California, Berkeley, CA 94720}
\affiliation{Miller Institute for Basic Research in Science, University of California, Berkeley, CA 94720}
\email{yongzheng@berkeley.edu}

\author[0000-0003-1455-8788]{Molly S.\ Peeples}
\affiliation{Space Telescope Science Institute, 3700 San Martin Drive, Baltimore, MD 21218}
\affiliation{Department of Physics \& Astronomy, Johns Hopkins University, 3400 N.\ Charles Street, Baltimore, MD 21218}

\author[0000-0002-2786-0348]{Brian W. O'Shea}
\affiliation{Department of Computational Mathematics, Science, and Engineering, 
Department of Physics and Astronomy, 
National Superconducting Cyclotron Laboratory,  
Michigan State University}

\author[0000-0002-6386-7299]{Raymond C. Simons}
\affiliation{Space Telescope Science Institute, 3700 San Martin Drive, Baltimore, MD 21218} 

\author[0000-0003-1785-8022]{Cassandra Lochhaas}
\affiliation{Space Telescope Science Institute, 3700 San Martin Drive, Baltimore, MD 21218} 

\author[0000-0002-0646-1540]{Lauren Corlies}
\affiliation{Rubin Observatory Project Office, 950 N. Cherry Ave., Tucson, AZ 85719}

\author[0000-0002-7982-412X]{Jason Tumlinson}
\affiliation{Space Telescope Science Institute, 3700 San Martin Drive, Baltimore, MD 21218}
\affiliation{Department of Physics \& Astronomy, Johns Hopkins University, 3400 N.\ Charles Street, Baltimore, MD 21218}

\author[0000-0002-6804-630X]{Britton D. Smith}
\affiliation{Royal Observatory, University of Edinburgh, United Kingdom}

\author[0000-0001-7472-3824]{Ramona Augustin}
\affiliation{Space Telescope Science Institute, 3700 San Martin Drive, Baltimore, MD 21218}

\begin{abstract}
\noindent 
The circumgalactic medium (CGM) of the Milky Way is mostly obscured by nearby gas in position-velocity space because we reside inside the Galaxy. Substantial biases exist in most studies on the Milky Way's CGM that focus on easier-to-detect high-velocity gas. With mock observations on a Milky-Way analog from the FOGGIE simulation, we investigate four observational biases related to the Milky Way's CGM. First, QSO absorption-line studies probe a limited amount of the CGM mass: only $35\%$ of the mass is at high Galactic latitudes $|b|>20\degree$, of which only half is moving at $|v_{\rm LSR}|\gtrsim100~\kms$. Second, the inflow rate ($\dot{M}$) of the cold gas observable in \HI\ 21cm is reduced by a factor of $\sim10$ as we switch from the local standard of rest to the galaxy's rest frame; meanwhile $\dot{M}$ of the cool and warm gas does not change significantly. Third, \OVI\ and \NV\ are promising ions to probe the Milky Way's outer CGM ($r\gtrsim$15 kpc), but \CIV\ may be less sensitive. Lastly, the scatter in ion column density is a factor of 2 higher if the CGM is observed from inside-out than from external views because of the gas radial density profile. Our work highlights that observations of the Milky Way's CGM, especially those using \HI\ 21cm and QSO absorption lines, are highly biased. We demonstrate that these biases can be quantified and calibrated through synthetic observations with simulated Milky-Way analogs. 
\end{abstract}


\section{Introduction} 
\label{sec:intro}

With an analogy borrowed from the Sun, \cite{spitzer56} proposed that the Milky Way be surrounded by a hot corona in pressure equilibrium with cold dense gas seen in \CaII\ and \NaI\ absorption lines \citep{munch61}. 
The hot corona, now commonly referred to as the ``gaseous halo" \citep*{putman12} or ``circumgalactic medium (CGM)" \citep*{tumlinson17}, has been broadly detected with ultraviolet (UV) absorption lines in galaxies in the local Universe as well as at higher redshifts. Here we loosely define the CGM as those diffuse gas beyond a galaxy's ISM but within the virial radius. The CGM is a massive reservoir filled with gas that is multiphase, clumpy, and with complex kinematics; and it is omnipresent in various types of galaxies, such as low-mass dwarfs, $L\sim L*$ galaxies, and massive quiescent galaxies (for a non-exhaustive list: \citealt{chen10, tumlinson11, bordoloi11, stocke13, werk14, bordoloi14, liang14, borthakur15, lehner15, johnson15, nielsen16, burchett16, heckman17, ho17, keeney17, prochaska17, johnson17, rubin18b, lopez18, lau18, berg18, chen18, smailagic18, martin19, zheng19b, zahedy19, berg19}). 

The Milky Way's CGM has been broadly studied with multi-wavelength observations, such as \HI\ 21cm, UV, and X-ray. X-ray studies find that hot gas ($T\approx2\times10^6$ K) in the Milky Way's CGM has a total mass of $M_{\rm hot, <250\ kpc}\sim10^{10-11}~\msun$ \citep[e.g., ][]{miller15, faerman17, bregman18}. Meanwhile, \HI\ 21cm and UV observations focus on studying high-velocity clouds (HVCs; $|v_{\rm LSR}\footnote{$v_{\rm LSR}$ means the line-of-sight velocity with respect to observers at the local standard of rest (LSR).}|\gtrsim$ 90--100 $\kms$) in the Milky Way's CGM at distances of $\sim$5--15 kpc \citep{schwarz95, wakker97, wakker01, vanWoerden04, wakker07, thom08b, wakker08}. Unlike extragalactic studies where only impact parameters and transverse velocities can be obtained, HVC observations have the advantage of measuring radial velocities and gas infall rates, which are critical for accessing the role of CGM in sustaining the Milky Way's star-formation activities \citep[e.g., ][]{wakker08, shull09, lehner11, putman12, richter17, fox19}. 

However, that we reside inside the Milky Way's ISM at an off-center location ($R_0=8.2$ kpc, Table \ref{tb:gal_properties}; see \citealt{blandhawthorn16}) leads to many observational biases yet to be constrained. In this work, we aim to address four biases plaguing the studies of the Milky Way's CGM with mock observations of a Milky Way analog selected from a cosmological hydrodynamical zoom simulation -- Figuring Out Gas \& Galaxies in Enzo (FOGGIE; \citealt{peeples19, corlies18}). Hereafter, we refer to this mock Milky Way and its related synthetic observations as the ``Mocky Way.''

First, physical properties of the cool-warm gas in the Milky Way's CGM are biased toward those measured for \HI\ and ionized HVCs \citep[e.g., ][]{sembach03, lehner11, lehner12, richter17, fox19}, because gas at lower velocities (i.e., $|\vlsr|\lesssim100\kms$) is heavily contaminated by nearby ISM. These HVC-based studies are likely to underestimate the Milky Way's CGM mass by $\sim50\%$ \citep{zheng15}. Meanwhile, only HVCs at $|b|\gtrsim20\degree$ are available as limited by the distribution of detectable QSO sightlines, causing the Milky Way's CGM at $|b|\lesssim20\degree$ largely unattended.  

Second, because we observe the Milky Way's CGM from the location of the Sun, gas radial velocities are measured toward the LSR. Their tangential velocities and true trajectories in the Milky Way's CGM are not available. Therefore, there exists a bias in calculating gas flow rates $\dot{M}\equiv Mv/d$, where $M$, $v$, and $d$ denote gas mass, velocity, and distance. With no information on gas tangential velocities, often the LSR velocities are corrected to the Galactic Standard of Rest frame (GSR) to account for the co-rotation of the Sun with the disk. Depending on how $v$ and $d$ values are adopted, $\dot{M}$ values differ among literature work and often are difficult to compare. The simplest method is to apply $\vlsr$ and $d_{\rm LSR}$ directly, resulting in $\dot{M}_{\rm in}$ values biased toward observers at the LSR \citep[e.g., ][]{wakker07, lehner11}. Other methods involve: (i) correcting $d_{\rm LSR}$ to distance toward the Galactic center (e.g., \citealt{richter17}); (ii) converting $\vlsr$ to $\vgsr$ to account for disk rotation \citep[e.g., ][]{fox19}; (iii) reconstructing true gas velocity vectors toward the Galactic center with assumptions on gas tangential velocities and 3-dimensional spatial distribution \citep[e.g., ][]{thom08b, putman12}. 


Third, UV studies of the Milky Way's CGM are biased toward nearby dense gas within $\sim$15 kpc, which accounts for the majority of absorption-line features in QSO and stellar spectra (e.g., \citealt{lehner11, lehner12}). The Milky Way's outer CGM ($\gtrsim$15 kpc) has remained largely unattended to except the Magellanic Clouds and their associated gaseous features \citep{putman02, stanimirovic08, nidever08, richter13, fox14}. A general reasoning for this omission is that the outer CGM has too low density to manifest itself in QSO absorption lines. However, \cite{zheng19} show that the Milky Way's outer CGM should be taken into account in order to explain the distribution of \SiIV\ column densities measured towards 132 QSOs across the sky (see also \citealt{qu19}). Lastly, because we observe the Milky Way's CGM from the Sun and are biased toward nearby gas, it remains difficult to directly compare the Milky Way's CGM to other $L\sim L*$ galaxy halos. 

In this work, we aim to address and quantify these four observational biases by analyzing a simulated Milky Way analog from FOGGIE \citep{peeples19, corlies18} with consistent observational setups. In \S\ref{sec:sims} we provide information on the FOGGIE simulation, including feedback recipe, code information, resolutions, etc. In \S\ref{sec:gal_property}, we show physical properties of the simulated halo, and compare them with the Milky Way's. In \S\ref{sec:mocky_way}, we generate synthetic observations by putting a mock observer inside the simulated galaxy and building a galactic coordinate system consistent with that of the Milky Way. We show our results in \S\ref{sec:all_bias} which quantifies the four observational biases inherent in the Milky Way's CGM studies. In \S\ref{sec:caveat_future}, we discuss advantages, caveats, and future directions of using FOGGIE simulations for Mocky Way analyses. Lastly, we summarize in \S\ref{sec:summary}.  

For consistency, throughout this work we will use the terminology as the following unless otherwise specified. (I) We define the CGM as those diffuse gas within the virial radius (approximated by $\rvir$) of a galaxy but beyond the galaxy's immediate ISM/disk region. (II) Following the definition in \citet*{tumlinson17}, {\it cold}, {\it cool}, {\it warm}, and {\it hot} gas are defined as those with temperature of $T<10^4$ K, $10^4\leq T<10^5$ K, $10^5\leq T<10^6$ K, and $T\geq10^6$ K, respectively. (III) We define {\it high-velocity} gas as that with velocity of $|v|>100~\kms$, and {\it low-velocity} gas with $|v|\leq100~\kms$, regardless of the rest frame in use (see point IV). Note that, this definition of low-velocity gas includes the intermediate-velocity ($30\lesssim|\vlsr|\lesssim100~\kms$) and low velocity ($|\vlsr|\lesssim30~\kms$) gas commonly used in the literature. (IV) We use the {\it galaxy's rest frame (GRF)} to identify velocity vectors pointing toward the galactic center. We emphasize that $v_{\rm GRF}$ is different from $\vgsr$ -- the latter means the gas velocity $\vlsr$ is corrected for the disk rotation, but its velocity vector is still pointed toward the LSR.

\section{Simulation Setup}
\label{sec:sims}
The simulation we analyze here is an updated version of the ``high resolution'' simulation of the Tempest halo first presented in \citet{peeples19} (hereafter \citetalias{peeples19}). We briefly summarize the simulation setup here, focusing on the changes made since \citetalias{peeples19}. 

In brief, the FOGGIE simulations are cosmological hydrodynamic simulations evolved with the block-structured adaptive mesh refinement code Enzo  (\citealp{bryan14}; last described in \citealt{brummel19}) using a flat $\Lambda$CDM cosmology ($1 - \Omega_{\Lambda} = \Omega_{\rm m} = 0.285$, $\Omega_{\rm b} =0.0461$, $h = 0.695$; \citealt{planck13}) and a Piecewise Parabolic Method (PPM) hydro solver. 
The Tempest halo was selected to have a Milky Way-mass at $z=0$  ($M_{\rm halo} \simeq 1.5 \times 10^{12}\,\msun$) and no significant merger (a mass ratio of 10:1 or lower) at $z<2$ \citep{simons20}.
The simulation domain is a $(100\,h^{-1}{\rm cMpc})^{3}$ box;  the Tempest initial conditions were generated using the ``cosmological zoom'' method, with a $256^3$ grid cell/particle base resolution and an effective resolution of $4{,}096^3$ particles (M$_{\rm dm} = 1.39 \times 10^6\,\msun$) in the region of interest, which is a Lagrangian region encompassing all of the particles within two virial radii of the galaxy at $z=0$. We then evolved these zoom simulations on the NASA Pleiades supercomputer with a maximum of 11 levels of adaptive mesh refinement (cell sizes of $\simeq 190$ pc/$h$ comoving) and 10 levels of forced (i.e, uniform) refinement (cell sizes of $\simeq 380$ pc/$h$ comoving) within a $\pm 100$ kpc/$h$ comoving box tracking the halo since $z=6$ in order to ensure resolution of small spatial scales within the CGM.

Throughout we consider the output at $z=0.102$ as this was the most recent output at the time of analysis. 
At $z=0.102$, the ISM resolution is 249 physical parsec and the CGM resolution within the refined region is 498 physical parsec. As described in \citetalias{peeples19}, the only feedback implemented is thermal feedback from supernovae.

We made two major changes to the setup presented in \citetalias{peeples19}. First, we decreased the minimum mass at which stars can form from $2\times 10^4\,\msun$ to $1000\,\msun$. This was motivated by finding that small halos at early times were not forming stars in the old simulation; the main impact of this change is an increase in early star and thus metal production. Second, we implemented self-shielding as a modification in the metagalactic UV background \citep{hm12} used by Grackle \citep{smith17} when calculating cooling rates; this implementation is described in more detail in \cite{emerick19}. As expected, this choice leads to higher \HI\ column densities, especially at early times. The ionization states of H and He are tracked natively in the simulation, whereas those of the metals are post-processed with the Trident tool \citep{hummels17}.

\begin{figure*}[t]
    \centering
    \includegraphics[width=0.95\textwidth]{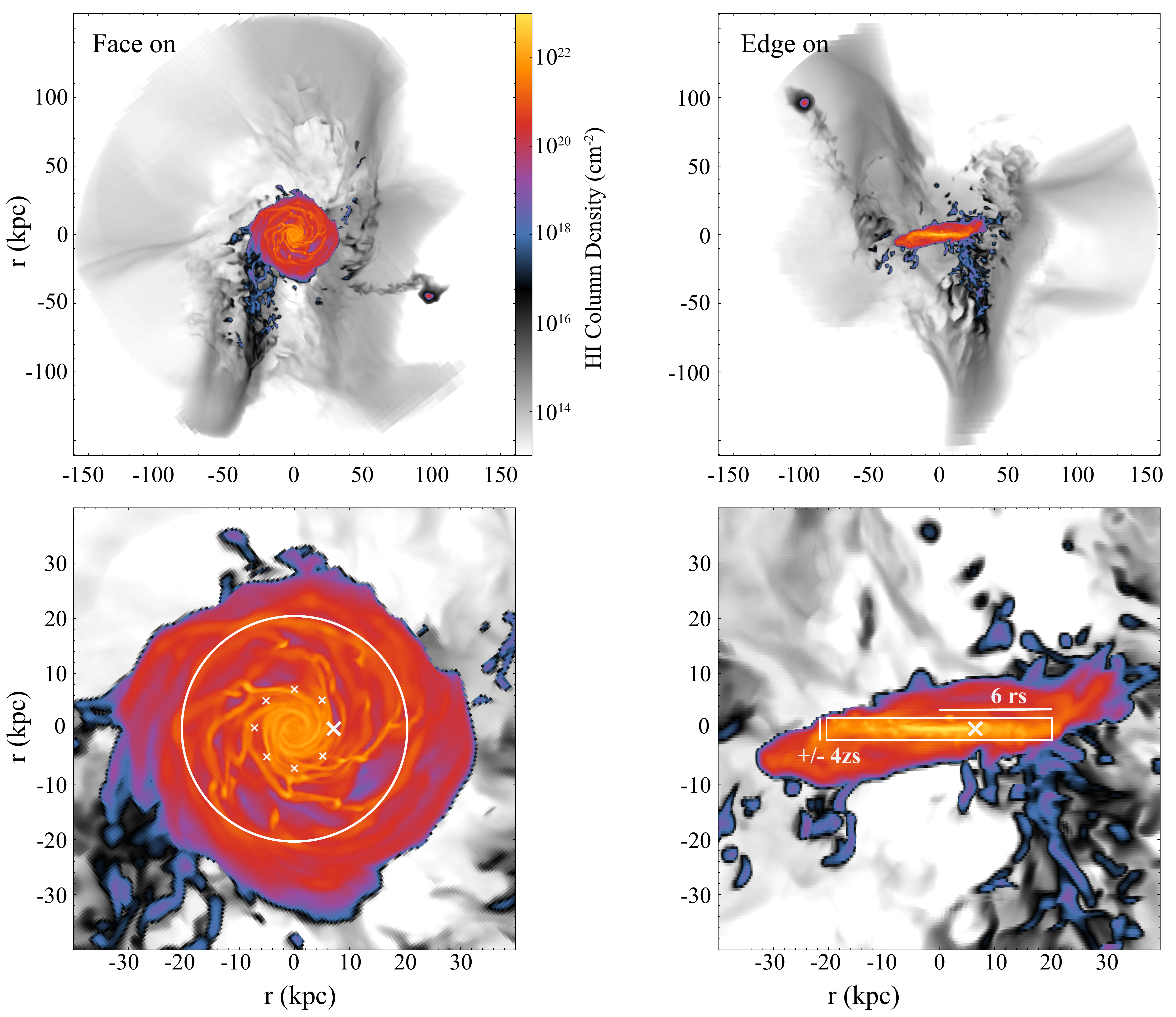}
    \caption{\HI\ column density of the Tempest halo at $z\simeq 0.1$ in face-on (left) and edge-on (right) projections. The top panels show \HI\ distribution within $\rvir$, and the bottom panels zoom into the inner 40 kpc to highlight the disk. The size of the disk is illustrated by a white circle in the bottom left panel and a rectangle in the bottom right panel, with disk radius of $6r_s$ and height of $\pm4z_s$ (see Figure \ref{fig:scale_r_z} for definition of $r_s$ and $z_s$). We place a fiducial off-center observer at 2$r_s$ in the disk plane (denoted with a thick cross) to be consistent with the position of the Sun in the Milky Way. We also select 7 other off-center locations (smaller crosses) 45$\degree$ apart from each other at the same solar circle, which are used to estimate how physical quantities vary due to the observer's location (see \S\ref{sec:all_bias}). The color scheme in this figure is designed such that blue roughly reflects the $\NHI$ detection limit of existing galaxy surveys at $\NHI\sim10^{17-18}$ cm$^{-2}$ or higher (e.g., LAB, \citealt{kalberla05}; AGES, \citealt{auld06}; CHILES, \citealt{fernandez16}; HI4PI, \citealt{hi4pi16}; GALFA-\HI, \citealt{peek18}). Any structures with dark-grey colors may not be detectable with current \HI\ 21cm sensitivity. }
    \label{fig:fig_galaxy}
\end{figure*}

\begin{deluxetable}{ccc}
\tablenum{1}
\tablecaption{Galaxy Properties \label{tb:gal_properties}}
\tablewidth{0pt}
\tablehead{
\colhead{Variable} & \colhead{FOGGIE/Tempest} & \colhead{Milky Way}
}
\startdata
$z$ & 0.102 & 0.0\\
$R_{200}^a$ (kpc) & 161.5 & 211.6$\pm$21.7 \\
$M_{200}^b$ ($\msun$) & 0.49$\times$10$^{12}$ & (1.1$\pm$0.3)$\times$10$^{12}$ \\
$M_{\rm bary, 200}^c$ ($\msun$) & 8.38$\times$10$^{10}$ & (8.5$\pm$1.3)$\times10^{10}$ \\ 
$M_{*, 200}^d$ ($\msun$) & 5.37$\times$10$^{10}$ & (5$\pm$1)$\times10^{10}$ \\
$R_0^e$ (kpc) & 6.8 & 8.2$\pm$0.1 \\ 
$r_{\rm s}^f$ (kpc)  & 3.4 & 2.74 -- 5.55 \\  
$z_{\rm s}^g$ (kpc)  & 0.5 & $0.08-0.22$ at $<$R$_0$ \\
$M_{\rm hot\ halo}$ ($\msun$) &  1.3$\times$10$^{8, h}$ & (2.8$\pm$0.5)$\times$10$^{10, i}$ \\ 
$M_{\rm HI\ HVC}$ ($\msun$) & 0.23$\times$10$^7$ (1.3\%)$^j$ & 2.6$\times$10$^{7, k}$\\
$M_{\rm ionized\ HVC}$ ($\msun$) & 0.16$\times$10$^8$ (46\%)$^l$ & 1.1$\times$10$^{8, m}$ \\ 
$M_{\rm CGM, gas}$ ($\msun$) & 1.19$\times$10$^{10, n}$ & N/A \\ 
\hline
\hline
\enddata
\tablecomments{
($a$) Virial radius; the Milky Way value is from \cite{blandhawthorn16} (hereafter, \citetalias{blandhawthorn16}). 
($b$) Total baryonic and dark matter mass within $\rvir$; the Milky Way value is a straight mean from \citetalias{blandhawthorn16}'s table 8.
($c$) Total baryonic (star+gas) mass within $\rvir$; the Milky Way value is from \citetalias{blandhawthorn16}. 
($d$) Total stellar mass within $\rvir$; the Milky Way is from \citetalias{blandhawthorn16}. 
($e$) Distance of the observer (the Sun) from the galactic center; the Milky Way value is a weighted-mean from \citetalias{blandhawthorn16}'s table 3. 
($f$) \HI\ disk scale radius; the Milky Way value is from \cite{diplas91, kalberla07, kalberla08}. 
($g$) Disk scale height; the Milky Way value has excluded the Galactic nuclear region \citep{lozinskaya63, dickey90, putman12}. 
($h$) Hot gas mass within $\rvir$ of the simulated CGM. 
($i$) Hot gas ($T\gtrsim10^6$ K) mass within 250 kpc \citep{bregman18}, estimated with a $\beta$-model density profile $n(r)\propto r^{-3/2}$ ($\beta\approx$0.5). 
($j$) Mass of the cold high-velocity gas in the CGM within 15 kpc, percentage in the parenthesis is the ratio of this component to the total cold CGM gas mass within 15 kpc. 
($k$) \cite{putman12}. 
($l$) Same as $j$, but for the cool and warm high-velocity CGM gas within 15 kpc. 
($m$) \cite{lehner11}. 
($n$) Total gas mass in the CGM. } 
\end{deluxetable}
 
\section{Properties of the Simulated Galaxy}
\label{sec:gal_property}

In this section, we discuss the physical properties of the Tempest  halo and galaxy at $z\simeq 0.1$, and compare them with those of the Milky Way (Table \ref{tb:gal_properties}). 

\begin{figure}[t]
\includegraphics[width=\columnwidth]{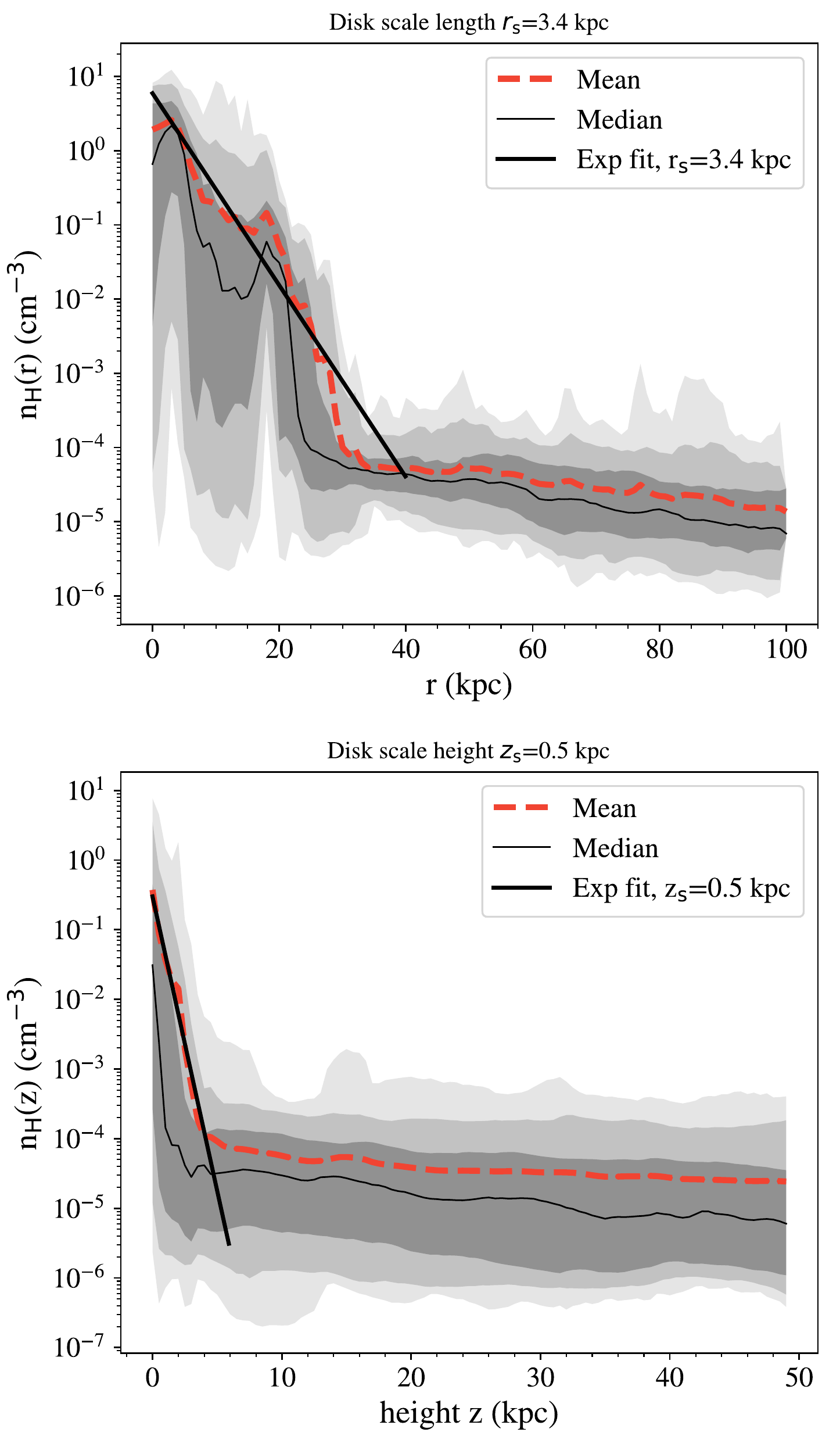}
\caption{\textbf{Top}: hydrogen (H) radial density profile. The red dashed line shows the volume-weighted mean $n_{\rm H}$ value and the thin black curve indicates the median value. The heavy to light grey bands show the $1\sigma$, $2\sigma$, and $3\sigma$ of the density distribution, respectively. We fit an exponential profile to the mean $n_{\rm H}(r)$ within 25 kpc and find a disk scale radius of $r_{\rm s}=3.4$ kpc, consistent with the Milky Way's value ($r_{\rm s}$=3.75 kpc; \citealt{kalberla08}). \textbf{Bottom}: $n_{\rm H}$ along the $z$ direction perpendicular to the disk. Similarly, we find a scale height of $z_{\rm s}=0.5$ kpc. Given the resolution of the simulation, we consider $r_{\rm s}$ well resolved but not $z_{\rm s}$ (see \S\ref{sec:mass_size_vrot}).} 
\label{fig:scale_r_z}
\end{figure}

\subsection{Mass, Size, and Rotation Curve}
\label{sec:mass_size_vrot}

In Figure \ref{fig:fig_galaxy}, we show the galaxy's \HI\ column density ($\NHI$) distribution in face-on and edge-on projections. The galaxy's virial radius ($\rvir=161.5$ kpc) is defined as the radius within which the mean density of the dark matter and baryons is 200 times the critical density of the Universe at $z=0.102$. At this simulation output, the galaxy has evolved to redshift $z=0.102$ with a well-settled gaseous disk. The edge-on projection shows a thin disk with a large warp extending beyond the disk. Although this work does not focus on the gaseous warp, we note that the galaxy is consistent with the Milky Way hosting a large extended warp detectable in \HI\ 21cm emission line \citep{diplas91, levine06, kalberla07, kalberla08}.

We define a cylindrical galactic disk at the center of the galaxy as shown in Figure \ref{fig:fig_galaxy}. The size of the disk is decided by examining the edge-on projection and make sure that the cylindrical region encloses the flat part of the thin gaseous disk. Quantitatively, the radius of the disk is taken as 6 times the disk's scale radius ($r_{\rm s}$; top panel, Figure \ref{fig:scale_r_z}) and its thickness $\pm4$ times the scale height ($z_{\rm s}$; bottom panel, Figure \ref{fig:scale_r_z}). Because we are interested in gas in the vast volume of the galaxy's CGM, the choice of the disk size does not affect our analyses and conclusion significantly. 

In Figure \ref{fig:scale_r_z}, we calculate $r_{\rm s}$ and $z_{\rm s}$ by fitting exponential functions $n(x)=n_0 e^{-x/x_{\rm s}}$ to the volume-weighted average radial and vertical hydrogen density profiles, where $x$ stands for $r$ (top panel) or $z$ (bottom panel). We find $r_{\rm s}=3.4$ kpc, consistent with the scale radius derived for the Milky Way's \HI\ disk (e.g., $r_{\rm s}= 2.74-5.55$ kpc by \citealt{diplas91}; $r_{\rm s}=3.75$ kpc by \citealt{kalberla08}). 
The scale height of the disk is $z_{\rm s}=0.5$ kpc. Because the cell size of the disk is 0.25 kpc, we consider the disk's scale radius well resolved but not the scale height $z_{\rm s}$, which is one of the caveats we discuss in \S\ref{sec:caveat_future}. It is noteworthy that the thickness of the Milky Way's \HI\ disk within the solar circle is 80--220 pc except near the Galactic nucleus region \citep{lozinskaya63, dickey90, putman12}, which means the simulation produces reasonable characteristic scale values at its current resolution.



In Table \ref{tb:gal_properties}, we compare some key properties of the simulated galaxy with those of the Milky Way. Within $\rvir$, we find that the total mass of the simulated galaxy is $M_{200}=0.49\times10^{12}~\msun$, including 83.0\% (by mass) of dark matter, 10.9\% of stars, and 6.1\% of gas. When compared to the Milky Way, the simulated galaxy is at the lower bound of the Milky Way literature values. For references, the total mass of the Milky Way within its virial radius is estimated to range from $5.5\times10^{11}~\msun$ to $2.6\times10^{12}~\msun$ \citep[e.g.][]{gibbons14, watkins10, watkins19} with a typical mean value of $(1.1\pm0.3)\times10^{12}~\msun$ \citep{blandhawthorn16}. When analyzing gas at different phases, we find that hot gas mass in the simulated CGM is two orders of magnitude lower than that estimated for the Milky Way \citep{miller15, bregman18}. Meanwhile, the masses of high-velocity cold gas (equivalent to \HI\ HVCs) and high-velocity cool-warm gas (equivalent to ionized HVCs) are both an order of magnitude lower than their corresponding observational values. Overall, we find that the simulated galaxy does not produce as much CGM mass as the Milky Way (see also \S\ref{sec:mocky_way} and \S\ref{sec:caveat_future}). 

\begin{figure}[t]
    \centering
    \includegraphics[width=\columnwidth]{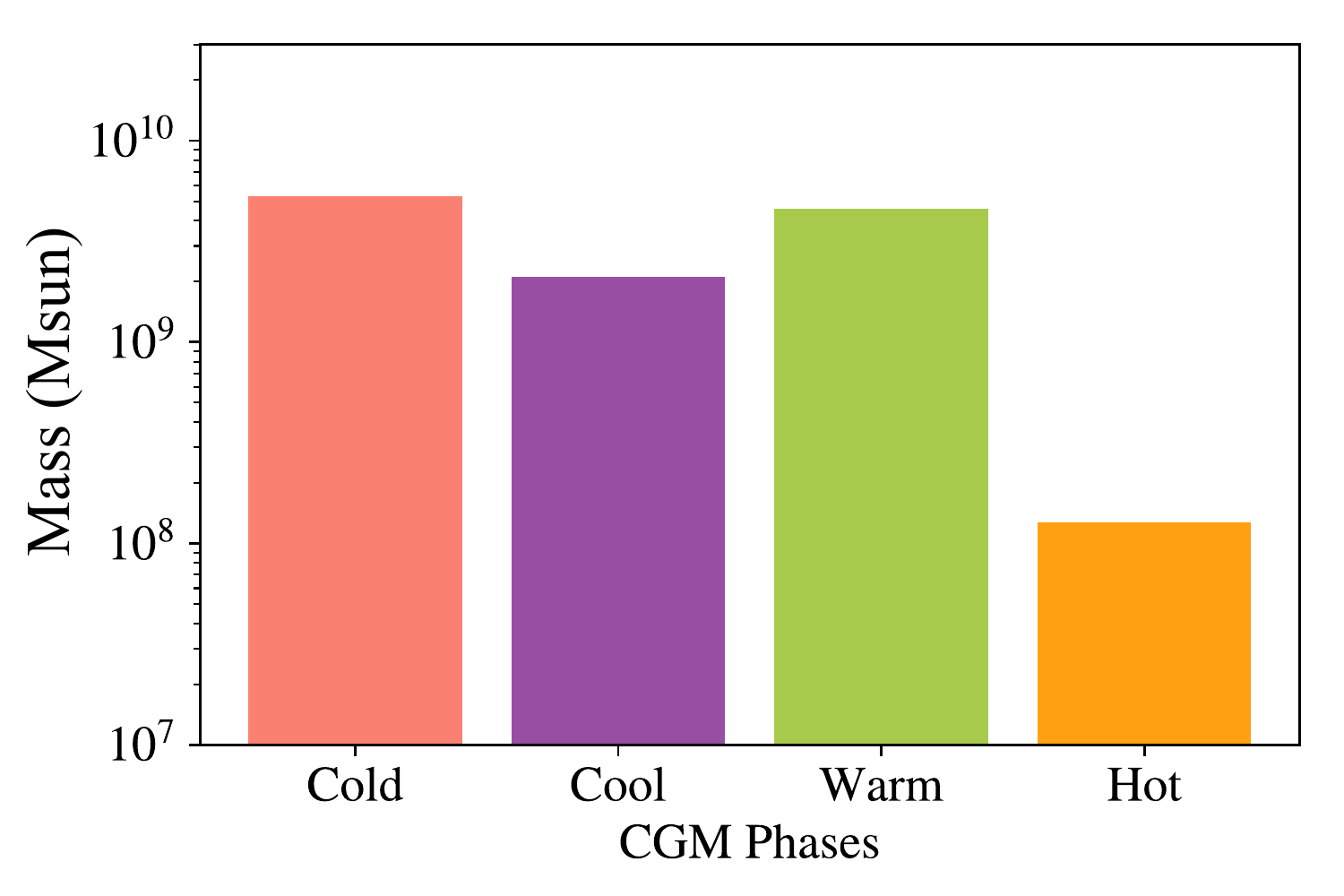}
    \caption{
    Masses of the cold ($5.2\times10^9~\msun$), cool ($2.1\times10^9~\msun$), warm ($4.5\times10^9~\msun$), and hot ($1.3\times10^8~\msun$) CGM gas within $\rvir$.  We find that the total gas mass in the cool and warm phases is a factor of two less than those estimated for $L\sim L^*$ galaxies at $z\sim$0.1--0.2. 
    } 
    \label{fig:cgm_mass}
\end{figure}

When compared to $L\sim L*$ galaxies at $z\sim$0.1--0.2, we find that the simulated galaxy is among the most abundant galaxy types and masses \citep[e.g., ][]{chen09, prochaska11a, werk13, tumlinson13, stocke13, borthakur15, burchett19}. In Figure \ref{fig:cgm_mass}, we split the simulated CGM gas into cold, cool, warm, and hot phases. As these masses are difficult to robustly constrain empirically, here we only aim for an order-of-magnitude comparison with observations to evaluate if the simulated CGM is representative for $L\sim L*$ galaxies. We find that the gas mass at the cool and warm phases is $\sim6.5\times10^9~\msun$, which is about a factor of 2 less massive than observational values ($\sim10^{10-11}~\msun$ for cool-warm CGM gas at $T\sim10^{4-5.5}$ K; e.g., \citealt{chen10, tumlinson11, werk14, lehner17, prochaska17, keeney17, bregman18, lehner20}).

\begin{figure}[t]
    \centering
    \includegraphics[width=\columnwidth]{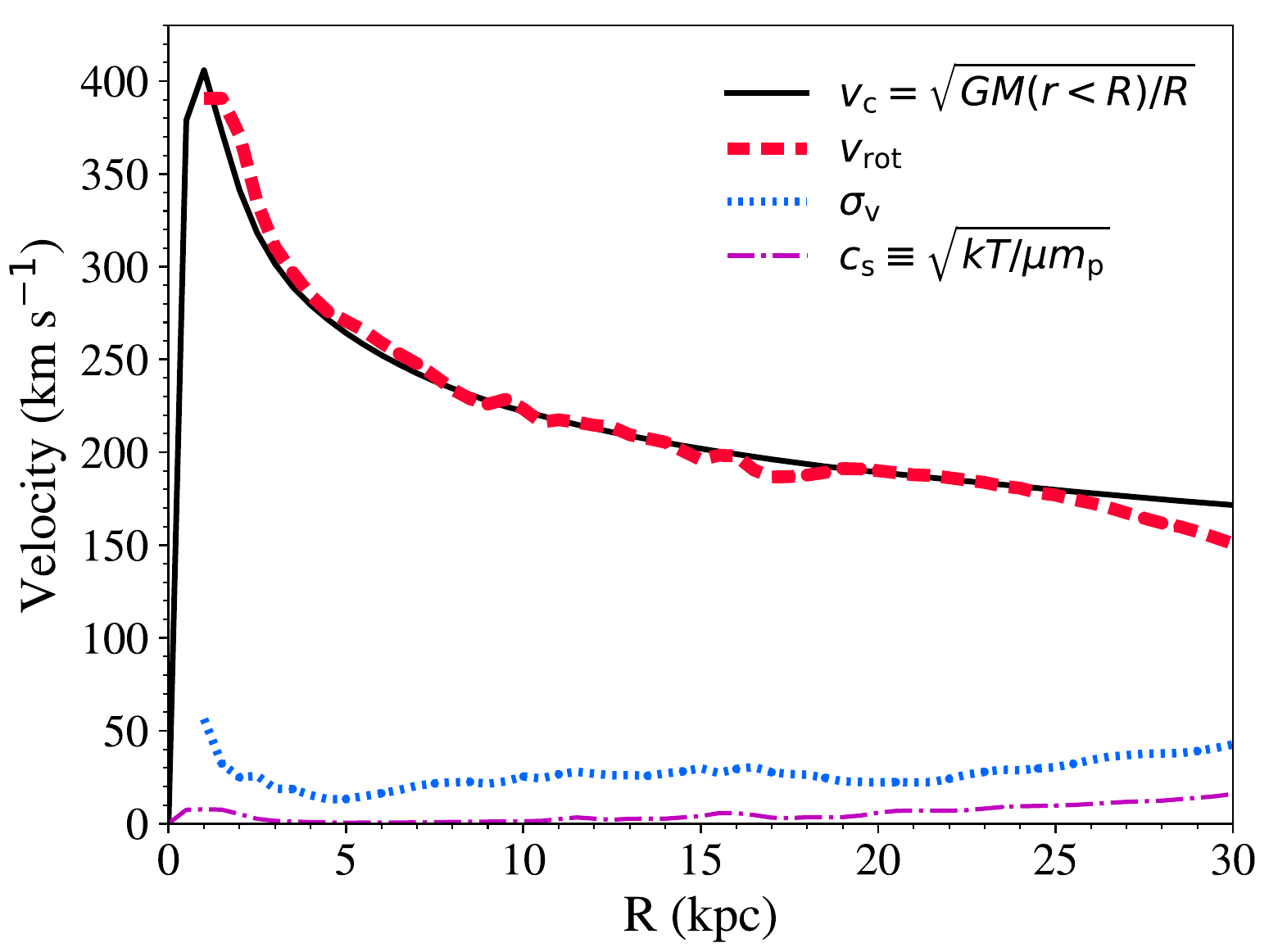}
    \caption{Circular velocity $v_{\rm c}$, mass-weighted gas rotation velocity $v_{\rm rot}$, velocity dispersion $\sigma_{\rm v}$, and mass-weighted sound speed $c_{s}$ of the simulated galaxy. As $v_{\rm rot}\approx v_{\rm c}$, the simulated galaxy is fully rotationally supported. $v_{\rm rot}$ is largely consistent with the Milky Way's rotation curve at $>5$ kpc. }
    \label{fig:vrot}
\end{figure}

Lastly, we check the rotation curve of the simulated galaxy in Figure \ref{fig:vrot}. The circular velocity is calculated as $v_c\equiv\sqrt{GM(<R)/R}$, where $M(<R)$ is the total dark matter and baryonic mass enclosed  within radius $R$. We follow the method described in \cite{el-badry18} and define the rotation velocity $v_{\rm rot}$ as the gas-mass-weighted mean azimuthal velocity $v_\phi$ per radial bin, where $v_\phi\equiv(x v_y - y v_x)/\sqrt{x^2 + y^2}$. In this calculation, $\vec{z}$ is along the angular momentum direction which is perpendicular to the disk plane, $\vec{x}$ is the direction from the galactic center to the fiducial off-center observer, and $\vec{y}$ follows the right-handed rule. The velocity dispersion $\sigma_v$ is mass-weighted standard deviation of $v_\phi$ values per radial bin. 
In general, $v_{\rm rot}$ matches $v_{c}$, suggesting that gas in the simulated galaxy is rotationally supported. Overall, the galaxy's $v_{\rm rot}$ is consistent with the Milky Way's rotation curve at $R\gtrsim5$ kpc (\citealt{blandhawthorn16}; also see \citealt{sofue09, koposov10, kafle12, reid14, lopez-corredoira14, kupper15}). 

\subsection{Phases, Kinematics, and Morphology} 
\label{sec:phase_vel_morph}

\begin{figure*}[t]
    \centering
    \includegraphics[width=0.95\textwidth]{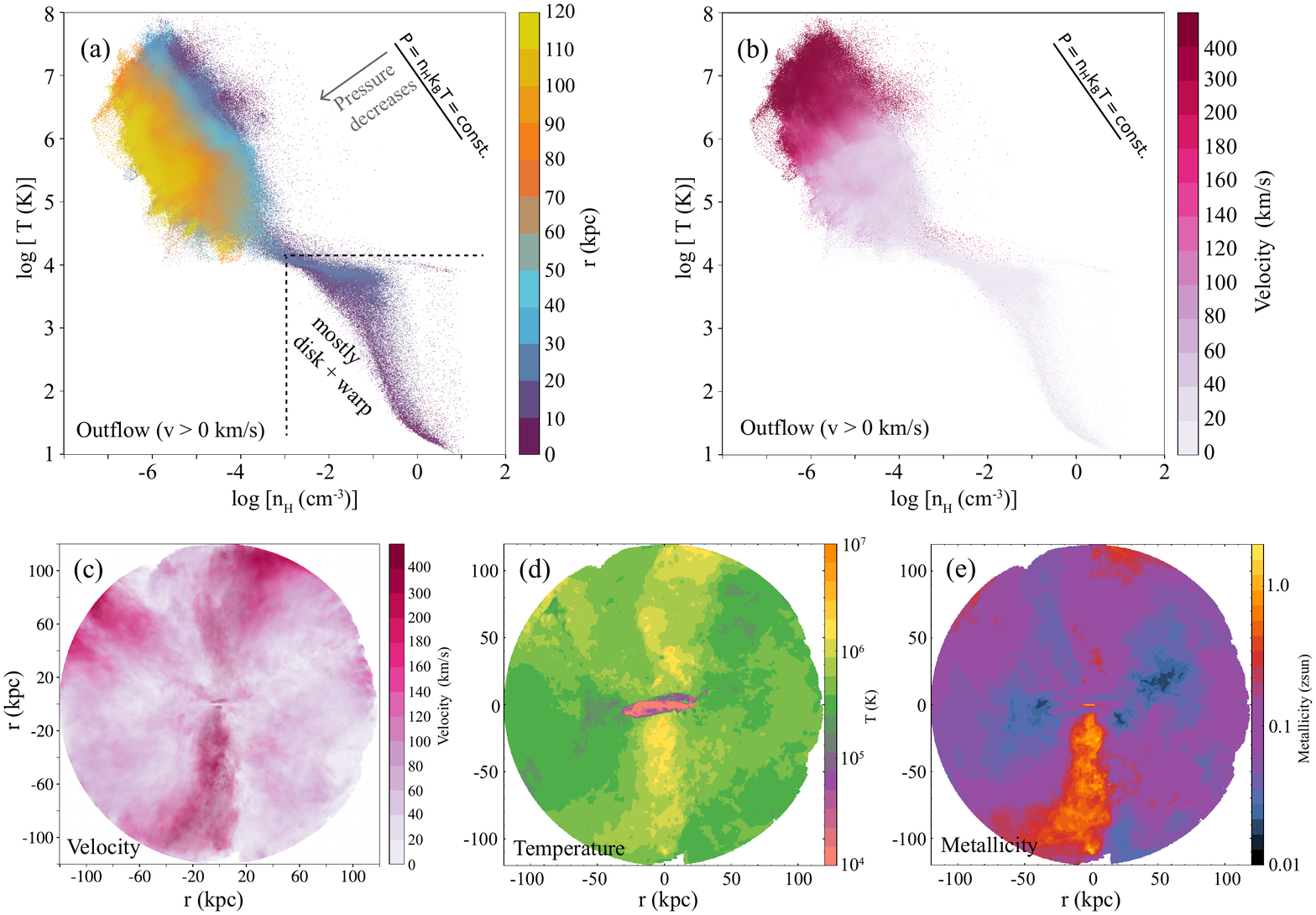}
    \caption{\textbf{Outflows} diagnosis ($v_{\rm r}>0~\kms$ in the GRF). Panels (a) and (b) are phase diagrams color-coded with gas radius and radial velocity, respectively. Panels (c), (d), and (e) show the CGM in edge-on projections color-coded with outflow radial velocity, density-weighted temperature, and density-weighted metallicity, respectively. In panel (a), thermal pressure gradually drops toward larger $r$ as indicated by a gray arrow. Note that the spread in pressure within each radius bin appears narrower because data points at larger $r$ are over-plotted on top of those at smaller $r$. Panel (b) shows that slower outflows are cooler and denser, regardless of the radius of the outflow. In panels (c-e), we find that slower outflows fill more volume with lower temperature and metallicity. Meanwhile, faster outflows appear to be bipolar on large scales, and they are identified with higher temperature ($T\gtrsim10^6$ K) and metallicity (Z from $\sim0.1$ to $>1~Z_\odot$).  The spatial and kinematic distribution of the outflows are in stark contrast with those of the inflows shown in Figure \ref{fig:phase_inflow}. }
    \label{fig:phase_outflow}
\end{figure*}

In this section, we outline the physical properties of the simulated CGM to form a baseline before proceeding with mock observations. We study gas inflows and outflows in the GRF with radial velocity $v_{\rm r}$ relative to the galactic center. We broadly define \textsl{inflows} as any CGM gas with $v_{\rm r}<0~\kms$, and \textsl{outflows} with $v_{\rm r}>0~\kms$, which likely include local motions such as turbulence and convection \citep{ford14}. We focus on the inflows' and outflows' bulk properties, such as phases, kinematics, spatial distributions, and morphology, but refrain from scrutinizing every interesting detail that arises. Investigations dedicated to analyzing the CGM's velocity structure (e.g., turbulence, convection) with FOGGIE simulations are forthcoming.

\begin{figure*}[t]
    \centering
    \includegraphics[width=0.95\textwidth]{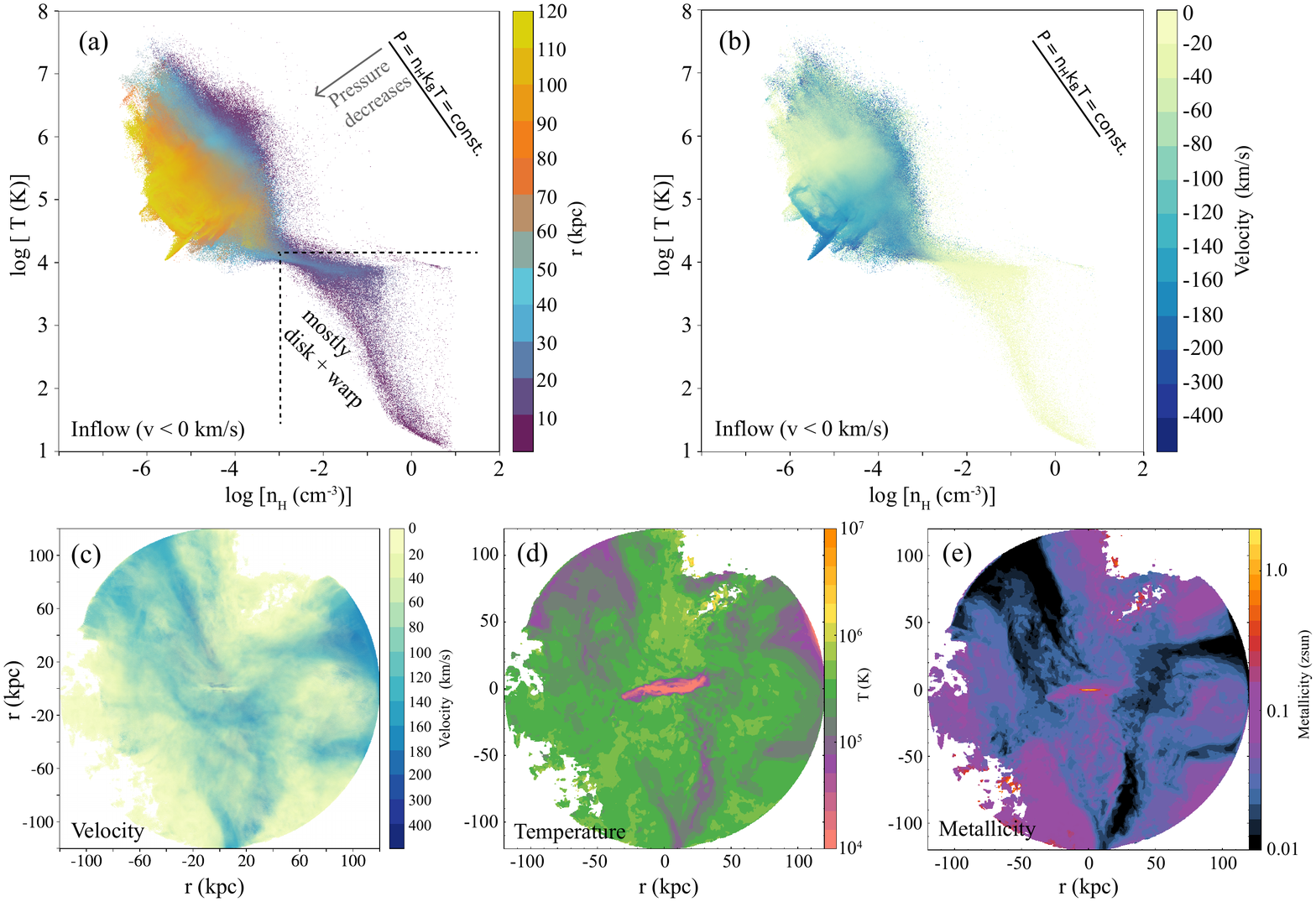}
    \caption{\textbf{Inflows} diagnosis ($v_{\rm r}<0~\kms$ in the GRF), with figure legends similar to Figure \ref{fig:phase_outflow}. In panel (a), we observe a similar radial pressure gradient in the phase diagram as that of the outflows. However, in panel (b), we do not find correlation between the inflow velocity and gas phase, different from what is observed for the outflows. Panels (c)-(e) show that fast inflows form large-scale filamentary structures, and they have low temperature ($T<10^{5}$ K) and metallicity ($Z\lesssim0.01~Z_\odot$). When compared to the outflows in Figure \ref{fig:phase_outflow}, we find that fast inflows bypass fast bipolar outflows. Meanwhile, slow inflows fill most of the volume of CGM and exist co-spatially with slow outflows, suggesting that these slow inflows and outflows may have different dynamical origins than their fast, large-scale counterparts. See \S\ref{sec:phase_vel_morph} for more detailed discussion. }
    \label{fig:phase_inflow}
\end{figure*}

In panels (a) and (b) of Figure \ref{fig:phase_outflow}, we show the outflows' phase diagrams color-coded with gas radius $r$ and radial velocity $v_{\rm r}$, respectively. Then in panels (c-e), edge-on projections of the CGM are illustrated with colors indicating the outflows' velocity, temperature, and metallicity, respectively. Cold dense ($T\lesssim10^4$ K, $n_{\rm H}\gtrsim10^{-3}$ cm$^{-3}$) gas seen in the lower right of panels (a) and (b) is mostly from the disk and warp, which we do not include in the following discussion. In panel (a), we find that within each 10 kpc radial bin, the outflows remain in thermal pressure equilibrium and the gas pressure drops toward larger radii, in broad agreement with \cite{voit19}'s radial pressure profile. We also observe a similar thermal pressure gradient in inflows, as shown in Figure \ref{fig:phase_inflow}. We note that, despite the similarity in gas phases and thermal pressure gradient, the inflows and outflows are not spatially well mixed. This can be easily seen by comparing the outflows and inflows' distribution in panel (c) in Figures \ref{fig:phase_outflow} and \ref{fig:phase_inflow}. Instead, the similarity raises a question: are the outflows and inflows in local pressure equilibrium? 

\cite{lochhaas19} find that a high-mass halo ($M_{\rm h}=10^{12}~\msun$; \citealt{fielding17}) is able to maintain thermal pressure equilibrium in most part of its CGM. However, in a low-mass halo ($M_{\rm h}=10^{11}~\msun$), thermal and non-thermal pressure (e.g., ram pressure and turbulence) are equally important in regulating the CGM's dynamical status. In their low-mass halo, a combination of the thermal and non-thermal pressure is still not enough to support against gravity, suggesting that the CGM is not in dynamical equilibrium. Given that the simulated galaxy we analyze here has a halo mass between their high-mass and low-mass cases, thermal pressure in our simulated CGM may play a major supporting role while non-thermal pressure is also non-trivial. We note that, because \cite{lochhaas19}'s analyses are based on spherically symmetric idealized CGM lacking cosmological accretion or mergers, their results may not be directly applicable to the FOGGIE cosmological simulations. Though the question of whether the CGM maintains local pressure equilibrium is beyond the scope of this paper, it merits a more detailed investigation and will be examined in forthcoming FOGGIE analyses.

We proceed with panel (b) in Figure \ref{fig:phase_outflow} to analyze the outflows' velocity gradient revealed in the phase diagram. Slower outflows are cooler and denser. We do not expect these slower outflows to come from material cooling out of hotter medium, in which scenario both cool and hot gas should move at similar velocities \citep{li14, thompson16, schneider18} and
thus no significant velocity gradient should exist in the phase diagram. In the edge-on projections shown in panel (c)--(e), We find that slow and fast outflows occupy different volume space. Fast outflows form large-scale bipolar structures, and rush out of the disk along perpendicular directions. They are metal-enriched ($Z\sim$ 0.1--1.0 $Z_{\odot}$) and hot ($T\gtrsim10^6$ K). Conversely, slower outflows ($0<v_r\lesssim60~\kms$) exist in a much bigger volume, but are cooler ($T\lesssim10^6$ K) and mostly metal-poor ($Z\lesssim0.1~Z_\odot$). 
These slow outflows may not be directly related to feedback material from the galaxy; instead, they may be associated with local motions, such as convection and turbulence, that result in positive velocity signs \citep{ford14}. We further note that outflows are distributed in bipolar manner along the galaxy's minor axis, consistent with galactic wind morphology observed in star-forming galaxies \citep[e.g., ][]{bland88, veilleux05}. And it resonates with the hypothesis that bipolar outflows produce excess in ion equivalent widths measured along a galaxy's minor axis \citep[e.g., ][]{bordoloi11, kacprzak12, lan18, martin19}. 

In Figure \ref{fig:phase_inflow}, we find that the inflows exhibit a similar radial pressure gradient in the phase diagram (panel a) as that of the outflows. Slow inflows are metal enriched and can be found in a large volume in the CGM. Meanwhile, fast inflows form large-scale filamentary structures, and are generally found at cooler phases except those near the disk/warp region. Their low metallicity ($Z\lesssim0.01~Z_\odot$) indicates that these fast inflows are relatively pristine material from the intergalactic medium. In panel (c), we find that as these fast inflows approach the disk-halo interface of the galaxy within $\sim10$ kpc, their gas temperature and metallicity become indistinguishable with ambient medium, suggesting that mixing due to turbulence possibly plays a role.

In all, we study the structures and kinematics of the outflows and inflows in the CGM in Figures \ref{fig:phase_outflow} and \ref{fig:phase_inflow}, the knowledge of which will be used to understand the observational biases in the Milky Way's CGM studies in \S\ref{sec:all_bias}. 


\section{Setup of Synthetic Observations}
\label{sec:mocky_way}


We build a mock galactic coordinate system consistent with that of the Milky Way; similar approaches are adopted by \cite{nuza14} and \cite{zheng15} to conduct synthetic observations of simulated Milky Way analogs. We place mock observers at eight off-center locations as defined in Figure \ref{fig:fig_galaxy}. The off-center locations are 6.8 kpc away from the galactic center, which is twice the disk scale radius as informed based on the location of the Sun in the Milky Way. Among the eight locations, we randomly choose one as the fiducial off-center location (thick white cross in Figure \ref{fig:fig_galaxy}) and refer to it as needed in the following. We define the mock galactic longitude $\ell$ and latitude $b$ in a consistent way as those of the Milky Way. Standing at an off-center location, the mock observer will find the galactic center at $(\ell=0\degree, b=0\degree)$ and the anti-galactic center at $(\ell=180\degree, b=0\degree)$. $\ell$ increases counter-clockwise when we look down from the north galactic pole, where north is defined as the direction of the disk's angular momentum. $b$ is positive above the galaxy plane and negative below. We calculate the bulk velocity of gas within 1 kpc of the mock observer and adopt it as the local standard of rest (LSR). 

\begin{figure}
    \centering
    \includegraphics[width=\columnwidth]{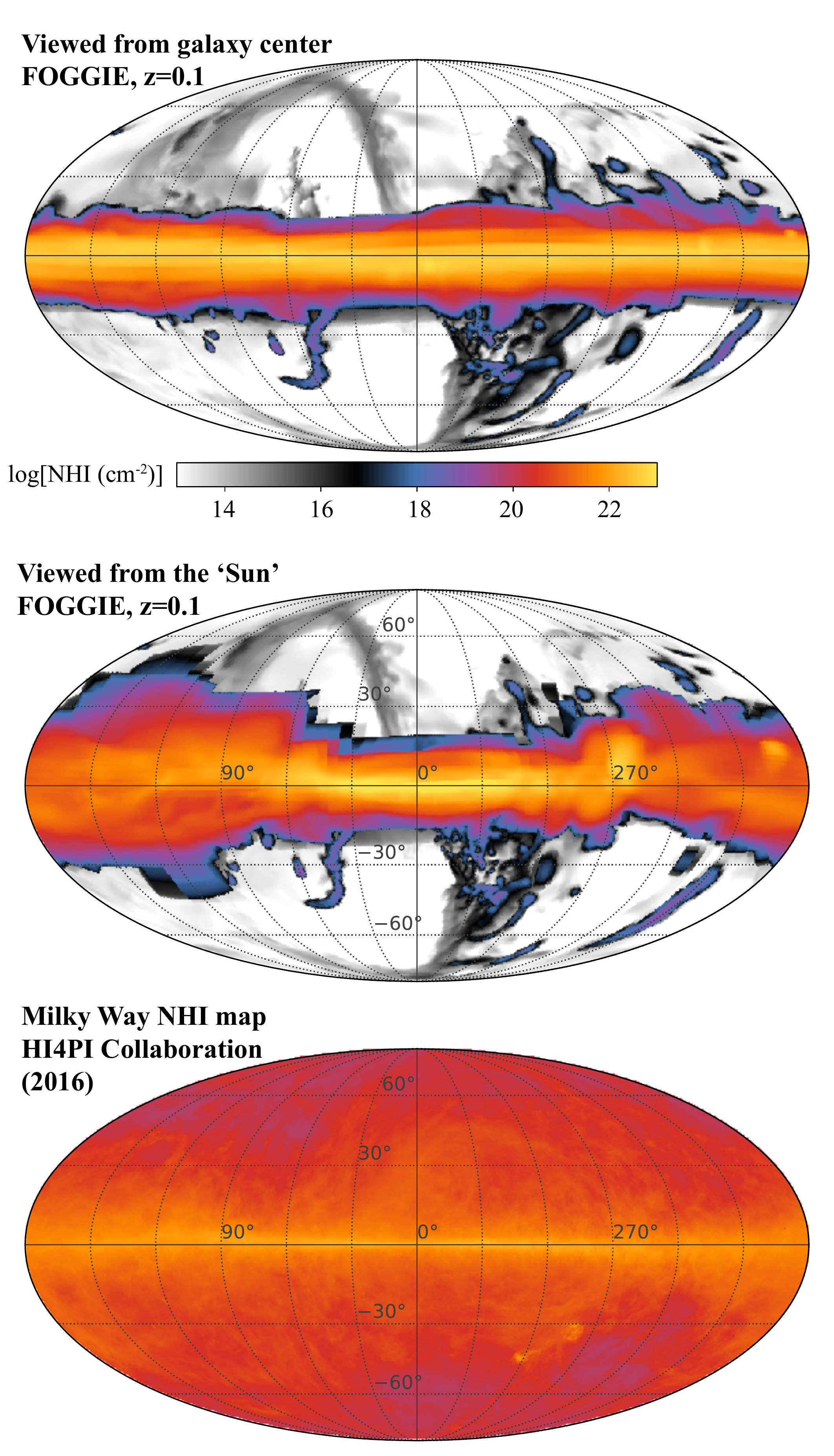}
    \caption{All-sky mollweide projection of $\NHI$. \textbf{Top}: $\NHI$ seen by an observer at the center of the simulated galaxy. All three maps share the same color bar. \textbf{Middle}: $\NHI$ seen by the fiducial observer placed at the off-center location consistent with where the Solar 
    System is in the Milky Way. \textbf{Bottom}: the Milky Way's \HI\ 21cm column densities measured within $\pm600~\kms$ of the LSR \citep{hi4pi16}. We find that: (1) at $|b|\gtrsim20\degree$, the simulated galaxy does not produce as much \HI\ as the Milky Way; (2) along the galactic plane at $|b|\lesssim20\degree$, the simulated galaxy exhibits similar column densities and gaseous structures as the Milky Way. Two reasons may be responsible for the discrepancy and similarity in the $\NHI$ all-sky distribution. First, the simulated halo is less massive than the Milky Way, resulting in lower \HI\ column densities. Second, the disk gravitational potential along the $z$ direction is not well resolved by FOGGIE, therefore, the gas distribution and column density above the galaxy's plane are incorrect. See \S\ref{sec:mocky_way} for more details.}
    \label{fig:nhi_allsky}
\end{figure}

In Figure \ref{fig:nhi_allsky}, we show \HI\ column density ($N_{\rm HI}$) in mollweide projections with the mock observer placed at the galactic center (top panel) and the fiducial off-center location (middle panel). We include \HI\ both in the disk and the CGM to compare with the Milky Way's observations. Results from the other 7 off-center locations are similar, with gas structures appearing at different $\ell$. Standing at the galactic center (top panel), the disk appears flat with small cloud clumps and filaments extending beyond the plane. However, when moving to the fiducial off-center location (middle panel), we find that both the simulated disk and the extraplanar \HI\ structures are reshaped. The disk remains relatively flat and thin within $\ell=\pm90\degree$ --- this transformation is due to the mock observer moving away from the galactic center. Toward the outskirts of the galaxy, the \HI\ disk becomes more fluffy with a large warp (see Figure \ref{fig:fig_galaxy}) and the mock observer's line of sight goes through less dense regions. Therefore, large flaring in \HI\ is observed at $|\ell|>90\degree$.

As a sanity check, we show the all-sky distribution of \HI\ in the Milky Way from the HI4PI survey \citep{hi4pi16} in the bottom panel. The $\NHI$ values are integrated from $\vlsr=-600~\kms$ to $600~\kms$. When comparing the middle and bottom panels, we find that the simulation is unable to produce as much \HI\ as the Milky Way at $|b|\gtrsim20\degree$. The low $\NHI$ values are likely to be due to the fact that the thickness of the galaxy's disk is not well resolved in the simulation (see \S\ref{sec:mass_size_vrot}), therefore the gravitation potential, thus the gas distribution, along the z-direction is incorrect. Furthermore, the total mass of the simulated halo is about a factor of two less than the Milky Way's value (Table~\ref{tb:gal_properties}), resulting in systematically lower \HI\ column densities. On the other hand, at $|b|\lesssim20\degree$ we find that the simulated galactic plane reaches a similar level of $N_{\rm HI}$ value as the Milky Way. The simulated disk starts to flare around $l=90\degree$ and $270\degree$ because of the projection of the warp, which is consistent with the Galactic warp observed in the Milky Way (\citealt{diplas91, levine06, kalberla07, kalberla08}). 

The direct comparison with the HI4PI data also reveals the limitation of current FOGGIE simulations (as well as other cosmological zoom simulations). The HI4PI map shows very detailed \HI\ structures all over the sky down to its resolution limit (16 arcmin; \citealt{hi4pi16}). Additionally, sharper \HI\ fiber structures \citep{clark14} and compact cloud clumps \citep[e.g., ][]{putman02, saul12} are revealed when higher resolution is made available by the GALFA-\HI\ data (4 arcmin; \citealt{peek18}). Conversely, the middle panel shows smooth \HI\ disk and warps, indicating the ISM in the simulated galaxy is not well resolved and not suitable for detailed analysis. Therefore, we only focus on the CGM of the simulation, which is defined as the region beyond the designated cylindrical disk and within the galaxy's virial radius.

\begin{figure*}[t]
\includegraphics[width=\textwidth]{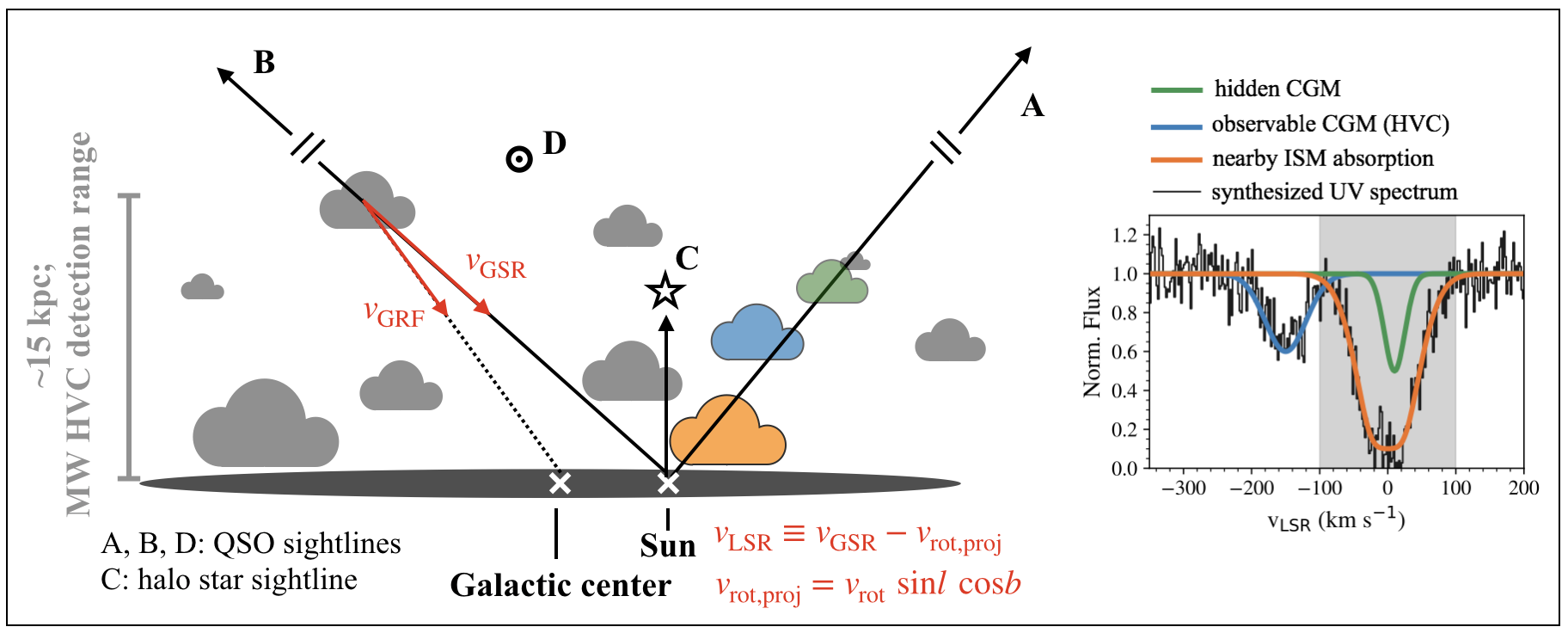}
\caption{Cartoon illustration of the four observational biases inherent in the Milky Way's CGM studies. \textbf{Bias I} (\S\ref{sec:bias_hvc}): toward sightline A, we revisit the question asked by \cite{zheng15} and estimate the fraction of the CGM mass omitted by QSO absorption-line studies. As shown in the inset spectral panel, only the high-velocity part of a UV spectrum is available to study the Milky Way's CGM. The low-velocity CGM gas is not accessible due to heavy contamination of the Milky Way's ISM. \textbf{Bias II} (\S\ref{sec:bias_restframe}): toward sightline B, we show that gas infall rate estimates for the Milky Way's CGM could be highly biased because gas velocities are measured relative to the Sun instead of the Galactic center. \textbf{Bias III} (\S\ref{sec:bias_stellar_qso}): combining sightline A (QSO) and C (halo star), we investigate whether QSO absorption lines are sensitive enough to probe the Milky Way's outer CGM beyond $\sim15$ kpc and which ions are the best tracers. \textbf{Bias IV} (\S\ref{sec:bias_in_out}): we look into the difference between the Milky Way's CGM (inside-out views; e.g., sightline A, B, C) and those of extragalactic systems (external views; e.g., sightline D), and investigate how the differences are related to the gas structures in the CGM. } 
\label{fig:bias_cartoon}
\end{figure*}

\section{Results: Quantifying the Observational Biases}
\label{sec:all_bias}

The purpose of this work is to identify and quantify four observational biases plaguing the studies of the Milky Way's CGM. In \S\ref{sec:intro}, we have briefly explained the four biases. Now we illustrate them in Figure \ref{fig:bias_cartoon} and address them in details in \S\ref{sec:bias_hvc}--\ref{sec:bias_in_out} with mock observations of the simulated FOGGIE halo. We caution that the statistics provided here is subject to the difference between the simulation and the Milky Way. Therefore, our result should be read as a qualitative assessment of the biases, with more work to be done in the future to systematically calibrate the biases.

\subsection{Observational Bias I: \\
How Much of the Milky Way's CGM is Omitted from QSO Absorption-line Studies?}
\label{sec:bias_hvc}

In the inset spectral panel in Figure \ref{fig:bias_cartoon}, we show an example UV line and highlight that only the high-velocity ($|\vlsr|\geq100~\kms$; white patch) part of the spectrum is available to the studies of the Milky Way's CGM. The low-velocity part (gray shade) is omitted because of heavy contamination from the Milky Way's ISM in the foreground. Even at the high-velocity regime, observers do not have full access to the Milky Way's CGM because most of the available QSO sightlines are located at $|b|\gtrsim20\degree$ \citep[e.g., ][]{sembach03, wakker12, lehner12}. Here we quantify how much gas in the Milky Way's CGM is omitted owing to data incompleteness in both the velocity and position space in QSO absorption-line studies. The result shown in this section is averaged over the eight off-center locations designated for the mock observer (see Figure \ref{fig:fig_galaxy}).

First, we split the simulated CGM into a high-$b$ region ($|b|>20\degree$) and a low-$b$ region ($|b|\leq20\degree$), and calculate the fraction of gas mass located in each region. On average, we find that the fraction of the CGM mass at the low-$b$ region is $f_{\rm lowb}\equiv M(|b|\leq20\degree)/M_{\rm all}\approx$ 65\%. When categorizing the simulated CGM gas into different phases, we find that 98\%, 49\%, 40\%, and 14\% of the cold, cool, warm, and hot gas mass is at the low-$b$ region, respectively. The fractional difference here suggests that the cold gas is preferentially found near the disk in a co-planar manner, whereas most of the hot gas occupies a volume perpendicular to the disk plane (see also panel d in Figures \ref{fig:phase_outflow} \& \ref{fig:phase_inflow}). For the cool and warm gas that are the most common to probe in UV, observers may have lost about half of the mass information of the Milky Way's CGM merely because of the lack of QSO sightlines at low Galactic latitudes.  

Meanwhile, not all of the CGM at the high-$b$ region is observable. Those moving at low velocity accounts for $f_{\rm lowV}\equiv M(|b|>20\degree, |\vlsr|<100~\kms)/M(|b|>20\degree)\approx49$\% of the CGM mass at the high-b region (see also \citealt{zheng15}). When breaking into different phases, the $f_{\rm lowV}$ values are 81\%, 30\%, 55\%, and 38\% for the cold, cool, warm, and hot gas, respectively.  


Overall, we find that the lack of QSOs at the low-$b$ region and the ISM contamination at the low velocity cause a severe loss of information of the Milky Way's CGM. In position and velocity space, only the CGM gas at $|b|\gtrsim20\degree$ and $|\vlsr|\gtrsim100~\kms$ is accessible, which accounts for $f_{\rm obs}=(1-f_{\rm lowb})(1-f_{\rm lowV})\approx18\%$ of the total CGM mass. And the $f_{\rm obs}$ values are $\sim0.4\%$, $36\%$, $27\%$, and $53\%$ for the cold, cool, warm, and hot gas, respectively. Qualitatively, this is consistent with \cite{zheng15}'s result where they find that the mass fraction of the observable CGM at $|b|\gtrsim20\degree$ and $|\vlsr|\gtrsim100~\kms$ is 35\% (see their Table 1). However, the exact $f_{\rm obs}$ value is subject to the details of each simulation. 



\begin{figure*}[t]
    \centering
    \includegraphics[width=\textwidth]{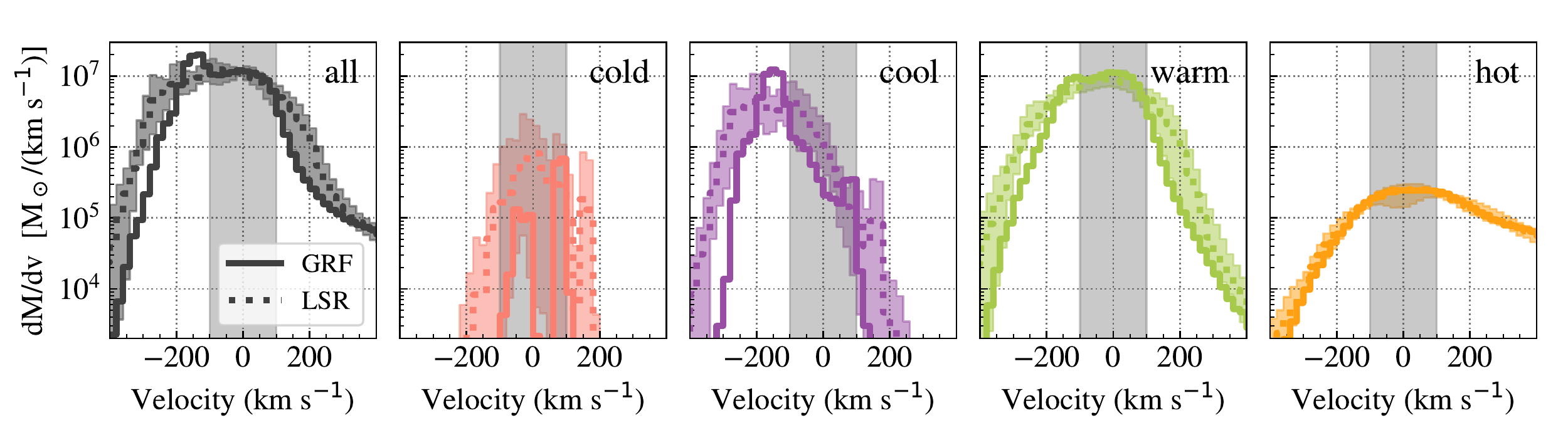}
    \caption{Gas mass as a function of velocity for all, cold, cool, warm, and hot CGM gas within $\rvir$. We only consider the CGM at $|b|>20\degree$ because the Milky Way's CGM is only accessible by high-$b$ QSO sightlines beyond the Galactic plane. The solid lines are values calculated in the GRF. The dotted lines and color shadings are calculated relative to the mock observer at the LSR: the dotted lines show values averaged over the eight off-center locations (see Figure \ref{fig:fig_galaxy}), and the color shadings denote the minimum and maximum values. The vertical grey shadings highlight the low-velocity spectral region often omitted in QSO absorption-line studies of the Milky Way's CGM. We find that: 
    (i) the cool and warm gas in the CGM are dominantly infalling regardless of the rest frame in use; (ii) when switching from the GRF to the LSR, the cool and warm gas masses are redistributed toward higher velocities as a result of the mock observer co-rotating with the disk at the LSR (see \S\ref{sec:bias_restframe}). }
    \label{fig:dmdv}
\end{figure*}

\subsection{Observational Bias II: \\Velocity Rest Frames:  $v_{\rm GRF}$, $v_{\rm GSR}$, and $v_{\rm LSR}$}
\label{sec:bias_restframe}



Direct measurements of gas flow rates in the rest frame of the Milky Way are unavailable because the gas velocity and distance are measured relative to the Sun instead of the Milky Way center. The sightline B in Figure \ref{fig:bias_cartoon} shows that the velocity vector thus mass flow rate are calculated with respect to local observers at the LSR. Even if we convert $\vlsr$ to $v_{\rm GSR}$ to correct for the disk rotation (e.g., as adopted in \citealt{fox19}), the velocity vector is still pointed toward the LSR, while the true velocity in the Milky Way's rest frame cannot be recovered owing to the ignorance of tangential motions of the gas perpendicular to the line of sight (see \citealt{thom08b, putman12}).

In Figure \ref{fig:dmdv}, we show how the CGM mass distribution with velocity changes as the mock observer moves from the off-center locations (i.e., the LSR; dotted line and shading) to the galaxy center (the GRF; solid line). We only consider the observable CGM gas at $|b|>20\degree$ as discussed in \S\ref{sec:bias_hvc}. We find that the masses of the cold, cool, and warm gas are more broadly distributed toward higher velocities in the LSR than in the GRF. The broader mass flux distribution with $\vlsr$ is caused by the mock observer's co-rotation with the disk at the LSR, which projects additional velocity components along the lines of sight \citep{wakker91}. 
Meanwhile, the mass flux distribution of the hot gas does not vary significantly between the GRF and the LSR. This is because the hot gas mostly exists at high galactic latitudes (see Figure \ref{fig:phase_outflow}), toward which the projected velocity of the disk rotation is trivial when switching the observing rest frames.

\begin{figure}
    \centering
    \includegraphics[width=0.9\columnwidth]{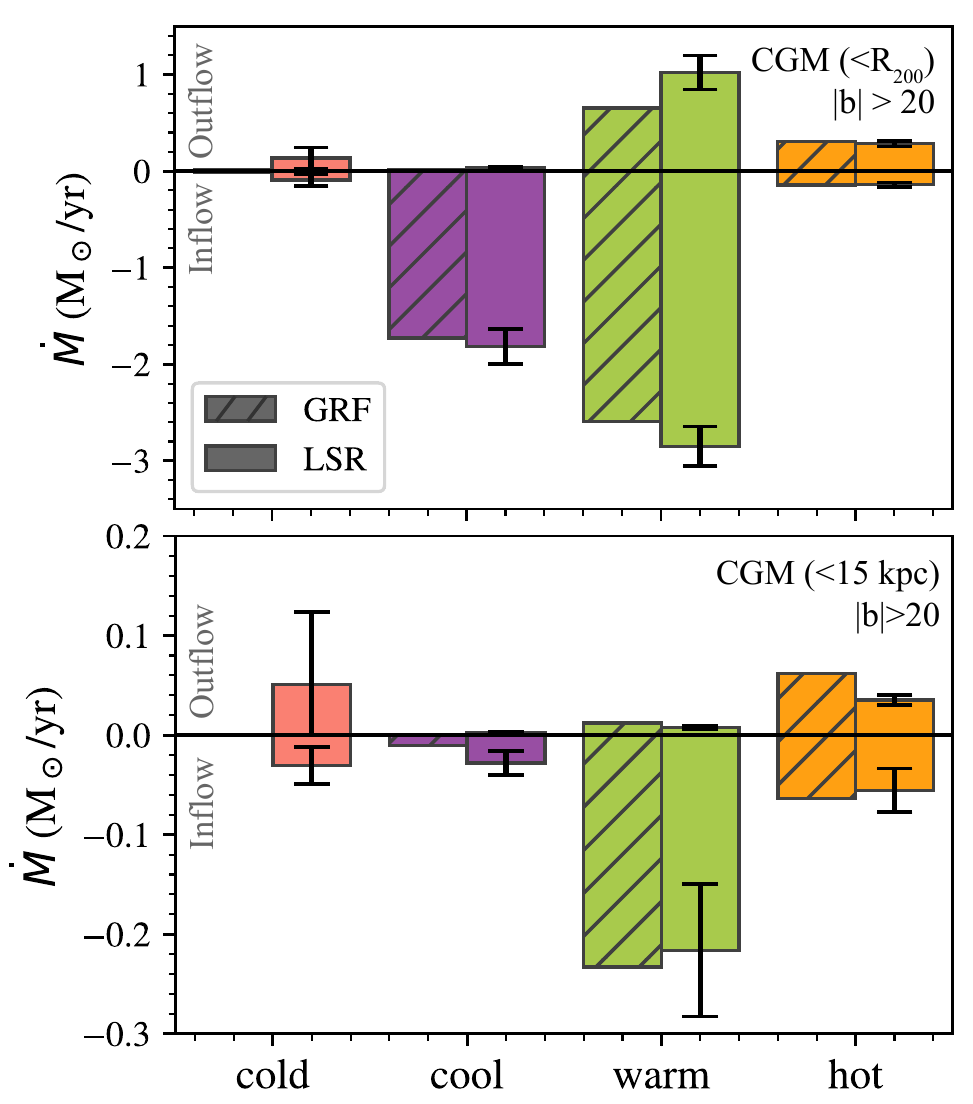}
    \caption{The inflow and outflow rates measured in the GRF (hatched) and the LSR (solid). Same as Figure \ref{fig:dmdv}, we only consider gas at $|b|>20\degree$. The top panel shows $\dot{M}$ for the CGM gas within $\rvir$, and the bottom one for gas within 15 kpc to compare with the Milky Way values. For $\dot{M}$ measured in the LSR, we show the mean and $1\sigma$ values of the eight off-center locations (see Figure \ref{fig:fig_galaxy}). 
    }
    \label{fig:dMdt_hist}
\end{figure}

We further compare the mass flow rates ($\dot{M}$) of the CGM gas measured in the GRF and the LSR. We define  $\dot{M}\equiv\Sigma M_i v_i/D_i$, 
where $M_i$, $v_i$, and $D_i$ denote the mass, velocity, and distance of a given gas cell toward the mock observer. This definition of $\dot{M}$ is widely used in the studies of gas inflows and ouflows \citep[e.g.,][]{lehner11, putman12, rubin14, zheng17, bordoloi17, fox19}. 
In Figure \ref{fig:dMdt_hist}, we show the inflow ($\dot{M}_{\rm in}\equiv\dot{M}<0$) and outflow ($\dot{M}_{\rm out}\equiv\dot{M}>0$) rates for the CGM gas within $\rvir$ (top panel) and for those within 15 kpc to compare with the Milky Way values (bottom panel). When measured in the GRF (hatched bars), $\dot{M}$ represents the true mass flow rate with respect to the galaxy center. Both panels show that the cool and warm gas at $|b|>20\degree$ are dominantly infalling. Within $\rvir$ (top panel), the net flow rate of the cool gas is $\dot{M}^{\rm GRF}=\dot{M}_{\rm in}^{\rm GRF}+\dot{M}_{\rm out}^{\rm GRF}\approx-1.7~\msunyr$ and that of the warm gas is $\approx-1.9~\msunyr$, which are consistent with what \cite{joung12} found for another simulated Milky Way analog. The net flow rate for the hot gas is $\approx0.16~\msunyr$, suggesting that most of the hot gas is outflowing. Its lower net flow rate compared to the cool and warm gas reflects that the hot gas accounts for less mass in the CGM as shown in Figure \ref{fig:cgm_mass}. Lastly, the net flow rate of the cold gas is $\approx0.01~\msunyr$, implying that within $\rvir$ the cold gas mass exchange between the CGM and disk is not significant.


When comparing the $\dot{M}$ values between the GRF (hatched) and the LSR (solid), we find that the inflow and outflow rates vary with the observing rest frames, although there is no consistent increase or decline in the $\dot{M}$ values from phase to phase. Here we focus on the inflow rates of the cold gas within 15 kpc (orange bars in the bottom panel), because in the Milky Way these values are relatively straightforward to derive from \HI\ 21cm data. In the LSR, the average inflow rate for the cold gas within 15 kpc is $\dot{M}_{\rm in}^{\rm LSR}\approx-0.03~\msunyr$. However, the inflow rate relative to the GRF is nearly a factor of 10 less; we find $\dot{M}_{\rm in}^{\rm GRF}\approx-0.004~\msun$, which is too small to show in the bottom panel. The low $\dot{M}_{\rm in}^{\rm GRF}$ value shows that the cold gas is mostly at rest with the disk. These gas is then shifted to higher $\vlsr$ velocity when observed in the LSR because of the co-rotation of the mock observer with the disk. When applied to the Milky Way, Figure \ref{fig:dMdt_hist} implies that the inflow rate of the \HI\ HVCs (i.e., the cold gas) is likely to be overestimated. For example, \cite{wakker07} estimate an infall rate of $\dot{M}_{\rm in}=$ -(0.1--0.25) $\msunyr$ for the \HI\ Complex C in the LSR, whereas \cite{putman12} suggest that the best-fit $\dot{M}_{\rm in}$ value for all the major \HI\ HVCs (including Complex C) is only -0.08 $\msunyr$ once the gas velocity is corrected from the LSR to the GRF with assumption on gas tangential velocity and 3-dimensional spatial distribution (see their section 6.1). 

Lastly, the inflow rates of the cool and warm gas within 15 kpc vary by less than a factor of 2 when we switch between the LSR and the GRF. This suggests that the $\dot{M}_{\rm in}$ values estimated for the ionized HVCs in the Milky Way's CGM \citep[e.g., ][]{lehner11, richter17, fox19} may not be significantly biased due to rest frame effects.

We note that the adopted definition for $\dot{M}$ does not reflect the actual infall rate because it assumes the gas would travel to the disk ($D_i$) at constant speed ($v_i$) without mass changes. Moreover, it assumes that the gas does not change its phase along the trajectory. Despite its broad use in observations, such a definition should be challenged because a number of simulations have shown that gas clouds travel in a galaxy's CGM may undergo phase and size changes due to interactions with ambient medium \citep{joung12, schneider18, gronke18, gronke19, banda-barragan19, li19}. Therefore, the $\dot{M}$ adopted here is also biased despite our effort to correct the observing rest frames. To further quantify the bias in $\dot{M}$, one needs to monitor the evolution of gas phases across time, which is beyond the scope of this work. We defer the corresponding analysis to another upcoming FOGGIE project. 


\begin{figure*}[t!]
    \centering
    \includegraphics[width=0.95\textwidth]{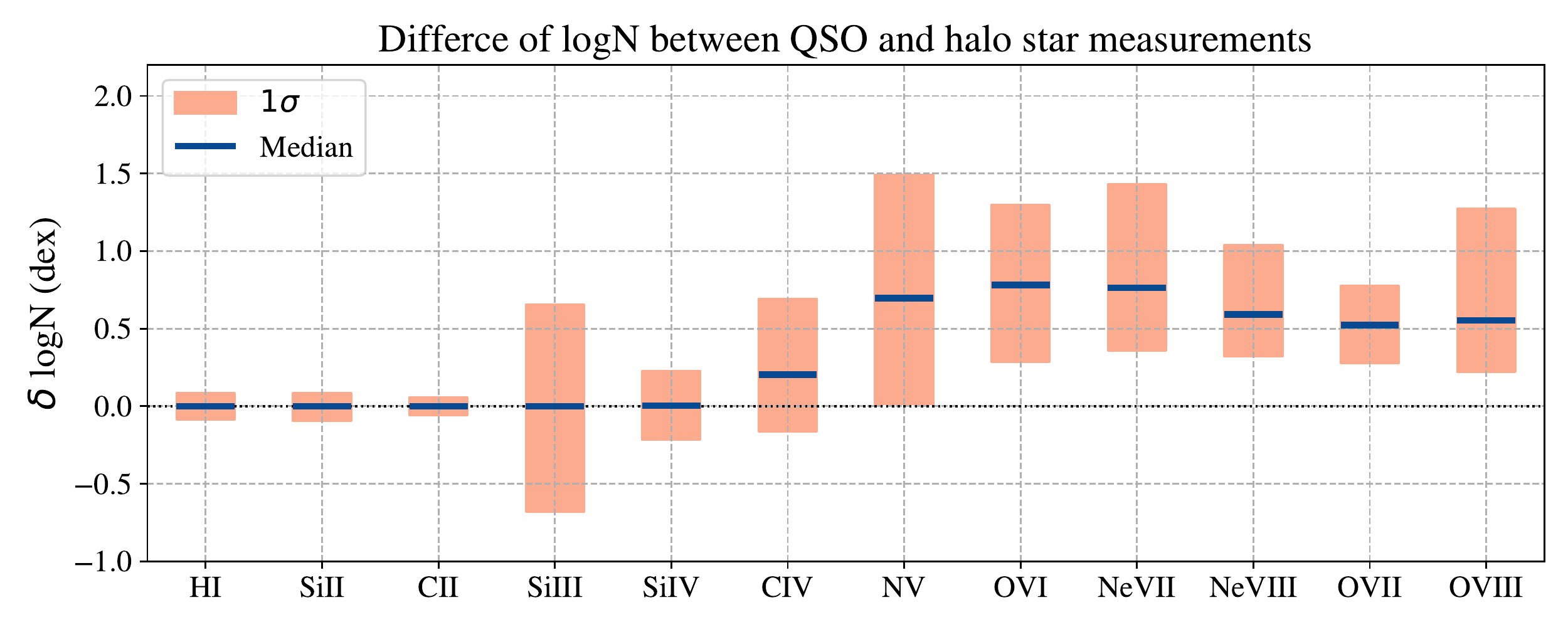}
    \caption{The statistics of $\delta\log N$ ($=\log N_{\rm qso}-\log N_{\rm star}$) based on a sample of 5000 random QSO-star pair sightlines with angular separations less than $10\degree$. We arrange the ions according to their ionization potentials. The mock stars are placed at 5-15 kpc and the QSOs at $\rvir$. All sightlines are located at $|b|>20\degree$. The dark-blue lines indicate the median values and the orange bars show $1\sigma$. If the simulated CGM were representative of the Milky Way's, this figure implies that: (1) it is more likely to detect the outer CGM ($>15$ kpc) using \NV\ and \OVI\ in UV or \NeVII, \NeVIII, \OVII, and \OVIII\ in X-ray through QSO-star pair experiments; and (2) \CIV, despite its positive median $\delta\log N$, shows ambiguous results about whether it can be used to trace the outer CGM given the large $1\sigma$ scatter due to the clumpiness of the CGM gas.}
    \label{fig:star_qso}
\end{figure*}

\subsection{Observational Bias III: \\
Stellar Sightlines v.s. QSO Sightlines}
\label{sec:bias_stellar_qso}


Commonly, QSO sightlines are treated the same way as distant halo stars to study the Milky Way's disk-halo interface that is within a few kpc of the Galactic plane \citep[e.g., ][]{savage09, wakker12}. However, it remains unclear whether the QSO sightlines bear additional information of the low-density gas in the Milky Way's outer CGM ($\gtrsim15$ kpc) given that every QSO sightline has to pass through it before reaching us. Therefore, there may exist a bias in studying the Milky Way's disk-halo interface if the outer CGM plays a non-negligible yet to-be-recognized role in QSO absorption lines.
In fact, \cite{zheng19} show that the \SiIV\ column densities measured along QSO sightlines across the Galactic sky cannot be fitted by a plane-parallel model built for the Milky Way's disk-halo interface \citep{savage09}. They propose that a global component resembling the outer CGM needs to be considered in addition to the plane-parallel geometry to better explain the all-sky \SiIV\ column densities \citep[see also, ][]{qu19}.

\begin{figure*}[t]
    \centering
    \includegraphics[width=0.9\textwidth]{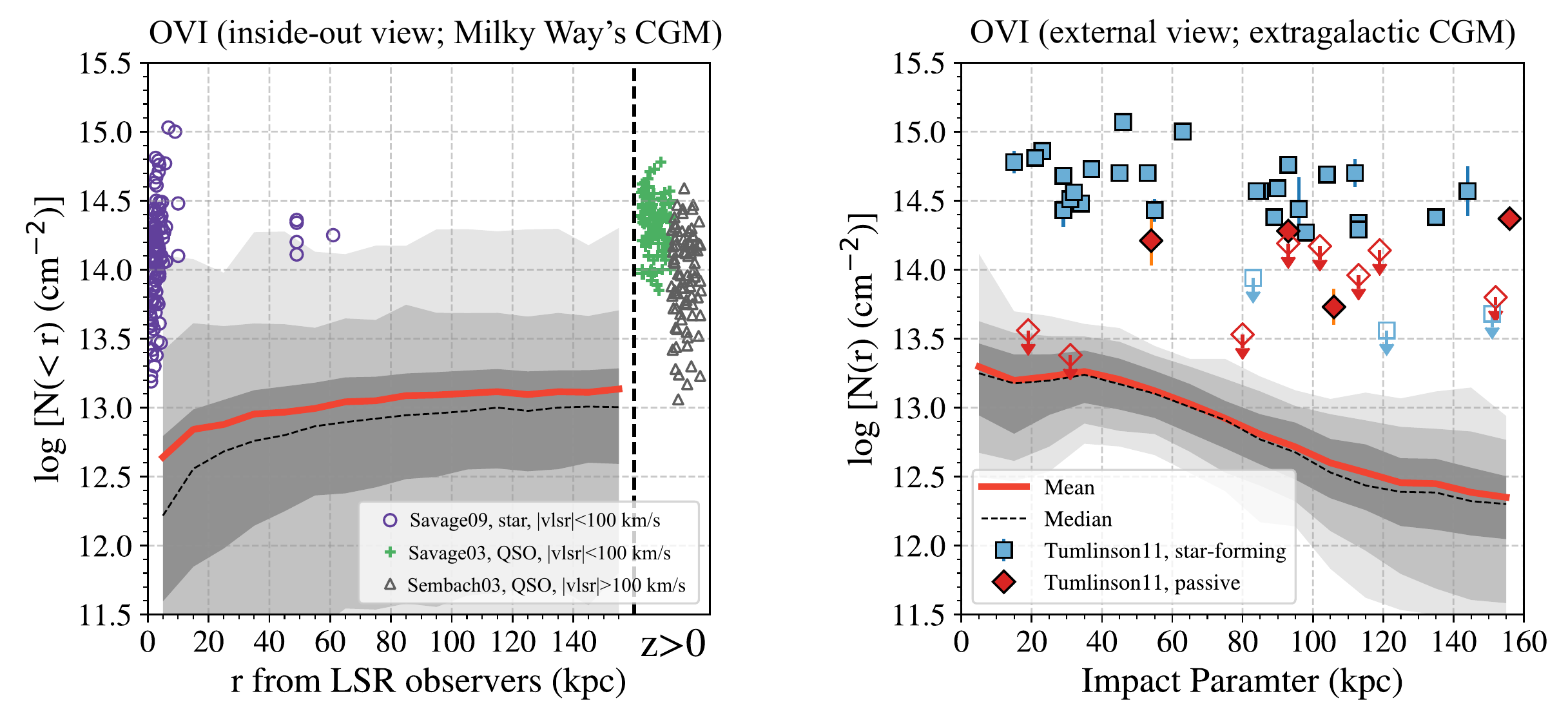}
    \caption{\textbf{Left}: the cumulative \OVI\ column density profile observed from inside-out views with $10^5$ random sightlines at $|b|>20\degree$. The red curve shows mean values and the gray bands indicate the $1\sigma$, $2\sigma$, and $3\sigma$ ranges. The open purple circles are the Milky Way measurements over the low-velocity range ($-100\lesssim \vlsr \lesssim 100~\kms$) toward halo stars with known distances \citep{savage09}. The green crosses are for low-velocity \OVI\ measured towards QSOs at $z>0$ \citep{savage03}, and the open gray triangles are for high-velocity gas at  $\vlsr\lesssim-100~\kms$ or $\vlsr\gtrsim+100~\kms$ \citep{sembach03}. Typical measurement errors for the datasets of \cite{savage09}, \cite{savage03}, and \cite{sembach03} are $\sim$0.1, 0.1, 0.08 dex, respectively. \textbf{Right}: $\log  \NOVI$($r$) profiles calculated with $10^5$ random sightlines through the simulated halo from external views. Each sightline has a path length of $2\rvir$ at a random impact parameter. The blue squares and red diamonds show \OVI\ measurements in star-forming and passive galaxies from the COS-Halos survey \citep{tumlinson11}. The figure shows that: (1) this simulation output under-produces $\log  \NOVI$, and (2) both the mock and real data show that the scatters in $\log  N$(\OVI) are much broader from the inside-out views (left) than from the external views (right).}
    \label{fig:in_ex}
\end{figure*}

Here we use the simulated CGM to evaluate whether the QSO absorption lines are sensitive to the low density gas beyond 15 kpc, and therefore address the potential bias in the studies of the Milky Way's disk-halo interface that treat QSO and halo star sightlines non-discriminately. Because the simulated halo is less massive than the nominal mass of the Milky Way (see Table \ref{tb:gal_properties}), we expect the mock observations to yield generally lower ion column densities than those observed (see also Figure \ref{fig:in_ex}). Therefore, we do not aim to compare the absolute column densities or ask if the column densities integrated only for the outer CGM is up to the sensitivity limit of current UV spectroscopy. Instead, we approach the question in a statistical and relative way by studying the variance of column density offset between the QSO and halo star sightlines. If a significant amount of low density gas exists in the outer CGM that is only available to the QSO sightlines, collectively there should be a positive column density offset between the QSO and halo star sightlines. Such an approach has been adopted by the QuaStar\footnote{\url{https://archive.stsci.edu/proposal_search.php?mission=hst&id=15656}} team (PI \citeauthor{peek18b}) to explore the Milky Way's outer CGM with pairs of QSO and halo star sightlines at close angular separations. Our mock observation, as described below, will complement QuaStar's experiment and provide insight to the interpretation of QSO-star pair comparison thanks to our prior knowledge of the simulated CGM's structure and kinematics (see \S\ref{sec:phase_vel_morph}).

In the QSO-star pair experiment, we place the mock observer at the fiducial off-center location. Note that the location of the observer in the disk does not matter significantly because we care about the gas beyond 15 kpc and no kinematic information is required. The mock stars are placed at 5--15 kpc from the observer, as informed by the current distance estimates of \HI\ and ionized HVCs in the Milky Way (see \S\ref{sec:intro}; and e.g., \citealt{wakker04, putman12, lehner12}). Meanwhile, the mock QSOs are placed at $\rvir$ to ensure that the whole volume of the CGM is well sampled. All the mock stars and QSOs are at $|b|>20\degree$ to mitigate the contamination from the ISM (see \S\ref{sec:bias_hvc}). We generate 5000 random QSO-star sightline pairs -- the maximum angular separation allowed for these pairs is set as 10$\degree$ as explained in the following.

Ideally, one would select a QSO and a star as close as possible, such that the difference between them traces the CGM gas beyond the star rather than the angular density variation in the foreground. However, because our mock observation is conducted on a grid-based simulation, each mock target is represented by a simulation cell with a non-negligible size. Without considering the spatial curvature, a simulation cell with size of 498 pc (see \S\ref{sec:sims}) is $2.9\degree$ wide if placed at 10 kpc. When placed in the mock galactic coordinate system ($b$ and $\ell$; as defined in \S\ref{sec:mocky_way}), the angular size of the cell ($\Delta b$, $\Delta \ell$) changes with the galactic latitude ($b$) of and the distance ($d$) to the cell as
\begin{equation}
    \Delta b \approx2.9\degree \left(\frac{\rm 10~kpc}{d}\right)~{\rm and}~ 
    \Delta \ell = \Delta b/{\rm cos} (b), 
\end{equation}
where ${\rm cos}(b)$ is to account for the fact that the radius of a latitude circle at $b$ shrinks by ${\rm cos}(b)$ compared to the great circle of the galactic equator. For example, a cell representing a mock star at ($d=10$ kpc, $b=20\degree$) has an angular size of $\Delta \ell\approx3.1\degree$ and $\Delta b\approx2.9\degree$; however, the same cell would appear to be stretched in the longitude direction with $\Delta \ell\approx33\degree$ if it is placed at $b=85\degree$. The same argument can be applied to pair sightlines represented by adjacent cells: despite that the cells are physically next to each other, their angular separations vary with their galactic coordinates, as constrained by the resolution of the simulation. Therefore, unlike the QuaStar experiment, there is no uniform value in the mock observation that can represent the angular separations of all possible close QSO-star pairs.

We decide that the maximum angular separation allowed for any pairs of mock QSOs and stars is 10$\degree$, in which case the pair sightlines do not go through more than 3 simulation cells in the foreground of the stars. By doing so, we limit the density variation in the foreground of the stars which may otherwise confuse the result of the QSO-star comparison. We note that changing the maximum angular separation to $20\degree$ or $30\degree$ does not affect our result significantly. For each pair of QSO-star sightlines, we calculate the logarithmic column density offsets ($\delta\log N=\log N_{\rm qso}- \log N_{\rm star}$), and show the median $\delta\log N$ and $1\sigma$ in values in Figure \ref{fig:star_qso}. A general impression is that $\delta \log N$ has a wide spread ($1\sigma\sim$0.3--0.8 dex) regardless of the ion considered, suggesting that the simulated CGM is highly clumpy for its rich gaseous structures (see Figures \ref{fig:phase_outflow} \&\ \ref{fig:phase_inflow}). Among all the ions, our result predicts that highly ionized species, such as \NV\ and \OVI, are better UV tracers of the outer CGM beyond 15 kpc for their positive $\delta\log N$ offsets. Weakly ionized species, including \SiII, \CII, \SiIII, and \SiIV, do not yield significant $\delta\log N$ for the outer CGM. It is unclear whether \CIV\ would be detectable in QSO-star pair sightline studies. The median $\delta \log N$ for \CIV\ is $\sim 0.2$\,dex, which means current spectroscopic instruments (e.g., HST/COS) should be sensitive to the \CIV\ absorption excess in QSO-star pair comparison. However, the clumpiness of the CGM from sightline to sightline may cause a larger $\log  N$ scatters ($1\sigma\sim$0.5 dex) and thus dominate over the $\delta\log N$ excess.

\begin{figure}[t]
    \centering
    \includegraphics[width=0.92\columnwidth]{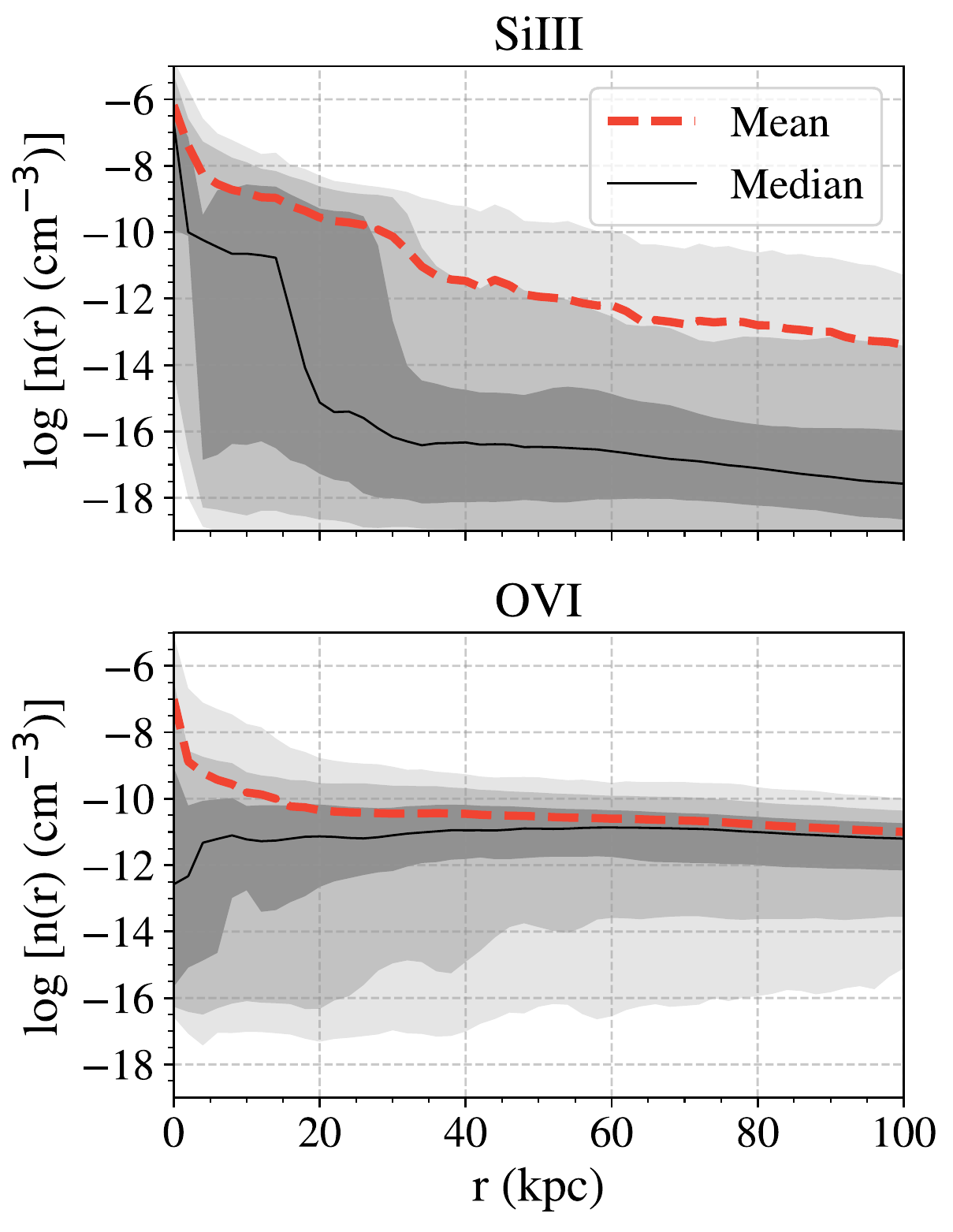}
    \caption{The radial density profile of \SiIII\ (top) and \OVI\ (bottom). The dark to light gray bands show $1\sigma$, $2\sigma$, $3\sigma$ values of the density distribution per radius bin. We find that the density of \SiIII\ quickly drops beyond 30 kpc, whereas the \OVI's remains relatively flat. The shapes of the ions' radial profiles determine their detectability in QSO-star pair experiments. See \S\ \ref{sec:bias_stellar_qso} and \S \ref{sec:bias_in_out}. }
    \label{fig:nr}
\end{figure}

We argue that the difference in $\delta\log N$ among various ions is rooted in their distinct radial density profiles. For example, as shown in Figure \ref{fig:nr}, \SiIII\ has a sharp decline in the $n(r)$ values around 30 kpc, whereas \OVI's $n(r)$ profile remains relatively flat out to large radii. Therefore, QSO sightlines will collect more \OVI\ from the outer CGM, enabling positive $\delta\log N$ values in Figure \ref{fig:star_qso}.  Generally speaking, highly ionized species are distributed more broadly in the outer CGM, whereas weakly ionized species tend to exist in denser environment closer to the disk. 

Existing and ongoing studies offer multiple ways to test our prediction. For example, similar levels of $\delta\log N$ scatters are reported in star-star pair sightline studies for a range of ions, including \OVI, \SiIV, \CIV, and \CaII\ \citep[e.g.,][]{howk02, bish19, werk19}. This implies that the clumpiness of the simulated CGM is approaching what is measured from observations, although we note that the QSO-star pairs in our experiments are limited by the resolution of the simulation. On the other hand, Figure \ref{fig:star_qso} suggests that \SiIV\ in the Milky Way's outer CGM may not be detectable, which is in stark contrast with \cite{zheng19}'s result; there \cite{zheng19} show that the Milky Way's outer CGM is traceable in \SiIV\ by analyzing $132$ QSO absorption lines across the Galactic sky. The discrepancy between the simulation and observation suggests that the simulated galaxy may not be identical to the real Milky Way, and our prediction of the detectability of the Milky Way's outer CGM in \CIV, \NV, and \OVI\ remains to be tested.


\subsection{Observational Bias IV: \\ Column Densities from the Inside vs.\ External Views}
\label{sec:bias_in_out}


Lastly, we tackle the conundrum that as the Milky Way residents, we always look at its CGM from the inside but observe extragalactic systems from external views. What is the observational bias when we compare the Milky Way's data points to others? As \cite{zheng15} alluded to, the path lengths of the Milky Way's CGM observations are only half of those in extragalactic systems at close impact parameters, therefore, the column densities measured from inside-out views are a factor of two lower on average. They suggest that the Milky Way's \OVI\ column densities should be consistent with extragalactic measurements \citep{tumlinson11} if taking into account the discrepancy in path lengths.

Here we discuss how the column density profiles and scatters differ between the inside-out and external views. In Figure \ref{fig:in_ex}, we show the \OVI\ profiles from the simulation and compare them with observations of the Milky Way's CGM \citep{sembach03, savage03, savage09} and extragalactic systems \citep{tumlinson11}. From both the inside-out and external views, we find that the simulated halo does not produce as much \OVI\ as the Milky Way and other $L\sim L*$ galaxies. When there are data available for other ions, we find a similar underproduction issue, such as \HI, \SiIV, \CIV, \NV, and \OVII\ (for the Milky Way observations, see \citealt{savage09, wakker12, fang15, zheng19, prochaska19}; for extragalactic observations, see \citealt{werk13, bordoloi14, liang14, borthakur15, lehner15, prochaska17, keeney17}). The lower mass of this FOGGIE halo may partially cause the underproduction, and we suspect that the thermal feedback recipe in FOGGIE may also be the culprit (see discussion in \S\ref{sec:caveat_future}). We note that the under-production of \OVI\ also occurs in other hydrodynamical simulations with various types of thermal feedback \citep[e.g., ][]{hummels13, ford16, liang16, gutcke17, suresh17, oppenheimer18b}. A promising solution is to implement non-thermal feedback (e.g., cosmic rays, \citealt{salem16, butsky18, ji19, kempski19}) that boosts the production of \OVI\ in the CGM (see reviews by \citealt*{somerville15, tumlinson17}).  



Nevertheless, both the mock and the real data show that the inside-out $\log \NOVI$ profile (left) has larger scatter than the external profile (right). Specifically, the inside-out observations of the Milky Way's CGM show $\log \NOVI$ scatters up to $\sim$1.5 dex, while the extragalactic data show $\lesssim$0.7 dex fluctuation if we only consider the CGM detection in star-forming galaxies. The difference between the inside-out and external views are related to the structures and thermal status of the CGM. First, the inside-out sightlines always have to pass through the inner CGM, which has higher density and is clumpier. In contrast, the external observations only pass through a small fraction of the inner CGM at close impact parameters, and most of the mock sightlines do not encounter large ion density variation beyond $r\gtrsim30$ kpc (see Figure \ref{fig:nr}). Second, as we show in Figures \ref{fig:phase_outflow} \& \ref{fig:phase_inflow}, bulk motions such as inflows and outflows occur radially. At a given radius, because of thermal pressure equilibrium, the hot gas is less dense than the cold gas. When such a CGM is observed from the inside, high $\log N$ scatters are measured as our lines of sight move from low-density hot outflow regions to high-density cold inflow streams. Conversely, when this CGM is observed from the external views, our lines of sight intercept the radial bulk motions tangentially. At a fixed impact parameter, the $\log N$ values are averaged over a wide range of radial features, which reduces the scatter in $\log N$.

\section{Advantages, Caveats, and Future Direction}
\label{sec:caveat_future}

The studies of the Milky Way's CGM have the advantage of directly measuring gas radial velocities; however, because the velocities are relative to the Sun instead of the Milky Way center, the interpretation of the Milky Way's CGM are biased. We offer a promising solution to quantify and calibrate the biases through synthetic observations of a simulated Milky Way analog. Many aspects of this work can be improved. Here we reflect upon the advantages and caveats in the FOGGIE/Mocky Way analysis, and discuss future direction for improvements. 

The FOGGIE cosmological simulation \citep{peeples19, corlies18} maps gaseous CGM structures at unprecedented spatial resolutions. It has been proven powerful to conduct zoom simulations to probe small scale structures and different phases and kinematics in the CGM \citep{vandevoort19, hummels19, suresh19, rhodin19}. In our Mocky Way analysis, the high, uniform resolution (0.19 kpc/$h$ comoving) in the CGM enables us to conduct synthetic observations such as QSO-star pair comparison, where the resolution in the inner halo becomes the limiting factor (see \S\ref{sec:bias_stellar_qso}). Moreover, the high resolution better resolves the density structure in the CGM \citepalias{peeples19}, which helps to quantify the column density scatter from sightline to sightline, and from inside and external views.

Caveats and challenges emerge as we progress. The first challenge is to select a Milky Way--like halo. By ``Milky Way--like", we mean consistency in mass, phase, kinematics, and gas structures. The current generation of the FOGGIE simulations focuses on six Milky Way-mass halos, which are described in detail in Paper IV \citep{simons20}. The simulation output used in this work is the least massive one among all the six FOGGIE halos. Despite its similar disk size ($r_s$, $z_s$), the simulation produces far less \HI\ and other ions in its CGM than the Milky Way (e.g., Figures \ref{fig:cgm_mass},  \ref{fig:nhi_allsky}, \ref{fig:in_ex}). As discussed in \S\ref{sec:mass_size_vrot} and \ref{sec:mocky_way}, the gravitational potential is not resolved in the $z$ direction, likely resulting in an incorrect distribution of \HI\ across the sky. Therefore, in future work, one should aim to increase the disk resolution to better simulate the structure of the ISM \citep[e.g.][]{christensen10}. Furthermore, higher resolution in and near the ISM will help us better understand how ions (e.g., \OVI\ and \SiIV) are distributed at the Milky Way's disk-halo interface \citep[][]{savage09}, which will further assist in separating out the contribution of these ions from the outer CGM. 

As we show in \S\ref{sec:bias_in_out}, FOGGIE's thermal feedback is unable to expel as many metals into the CGM and thus reproduce the high ion column densities as observed in the Milky Way and extragalactic CGM (see also \citealt{hamilton-Campos20}). Non-thermal feedback, such as radiation pressure from hot stars \citep{hopkins14} and magnetic fields and cosmic rays \citep{salem16, butsky18, ji19, kempski19}, is likely to be critical to alleviating these issues. 


Lastly, our mock observations do not consider how physical quantities are translated into observables through the lens of instruments. For example, we calculate the inflow and outflow rates by directly integrating the quantities in the velocity and position space. Ideally, synthetic spectroscopy (e.g., as in \citetalias{peeples19}) should be used to evaluate how the physical information may be missed because of systemic instrument effects. Our ultimate goal is to conduct synthetic spectroscopy to generate ion absorption lines and take into account observational limitation, such as instruments' point spread function, signal-to-noise ratio, Voigt-profile fitting, and sightline selection effect.

\section{Summary of the Mocky Way} 
\label{sec:summary}

Observations of the Milky Way's CGM are highly biased because we reside inside the Galaxy at an off-center location. The low-density gas in the CGM, especially those at low velocity, is mostly obscured by the ISM gas in the foreground. And, gas velocities are measured relative to the Sun instead of the Milky Way center, which affects the estimates of the gas inflow and outflow rates. In this work we investigate four observational biases related to the studies of the Milky Way's CGM. 

We conduct mock observations on a Milky Way analog from the FOGGIE simulation. The physical properties of the galaxy are largely consistent with the Milky Way values (see Table \ref{tb:gal_properties} and \S\ref{sec:gal_property}). The simulated CGM exhibits rich kinematic structures as shown in Figures \ref{fig:phase_outflow} and \ref{fig:phase_inflow}. We find bipolar outflows at high velocities ($v\gtrsim100~\kms$), high temperature ($T\gtrsim10^6$ K), and high metallicities ($\sim0.1$ to $>1 Z_\odot$). By contrast, large-scale filamentary inflows bypass the outflows;
they are observed to be moving at $v\lesssim-100~\kms$, with low temperature ($T\lesssim10^5$ K) and low metallicity ($Z\lesssim$0.01 $Z_\odot$). These filamentary inflows do not seem to be directly accreted onto the galaxy. Instead, they mix with ambient medium in the inner CGM at $r\lesssim20$ kpc. Low-velocity inflows and outflows fill most of the volume of the CGM. Overall, the cool and warm gas in the CGM are dominantly infalling toward the center of the galaxy, with a total net infall rate of $-3.6~\msunyr$ (Figure \ref{fig:dMdt_hist}). 

We build a mock galactic coordinate system similar to that of the Milky Way (\S\ref{sec:mocky_way}). The mock observer is placed at twice the disk scale radius to be consistent with the location of the Sun. In \S\ref{sec:all_bias}, we investigate four observational biases pertaining to the Milky Way's CGM studies, which are summarized as the following. 

\textbf{Bias I} (\S\ref{sec:bias_hvc}): we estimate how much mass in the Milky Way's CGM is omitted in QSO absorption-line studies that favor high-latitude ($|b|>20\degree$) and high-velocity ($|\vlsr|\geq100~\kms$) data. We find that the gas mass at $|b|>20\degree$ only accounts for 35\% of the total CGM mass, which means observations of the Milky Way's CGM may miss 65\% of the mass because QSOs are difficult to identify, let alone observe, at low Galactic latitudes. At $|b|>20\degree$, the high-velocity gas accounts for 51\% of the mass, suggesting that half of the CGM mass at high-$b$ is further omitted. In all, the observable mass of the Milky Way's CGM in position and velocity space is likely to be only 18\%.  
    
\textbf{Bias II} (\S\ref{sec:bias_restframe}): we calculate the mass flow rates $\dot{M}$ for the cold, cool, warm, and hot gas in the CGM at $|b|>20\degree$, and discuss how current inflow rate $\dot{M}_{\rm in}$ estimated for the Milky Way's CGM could be biased because the gas velocity is measured relative to the Sun (i.e., $\vlsr$ or $v_{\rm GSR}$) instead of the Milky Way center ($v_{\rm GRF}$). We find that the inflow rate of the cold gas within 15 kpc is reduced by a factor of $\sim 10$ when the gas velocity is corrected from the LSR to the GRF (Figure \ref{fig:dMdt_hist}). This suggests that the inflow rate of the Milky Way's \HI\ HVCs could be overestimated if the rest frame bias is not taken into account. Meanwhile, the inflow rates for the cool and warm gas change by less than a factor of 2 when switching between the LSR and GRF.

\textbf{Bias III} (\S\ref{sec:bias_stellar_qso}): we discuss the bias related to the studies of the Milky Way's disk-halo interface where QSO sightlines are not distinguished from halo star sightlines. We investigate whether QSO absorption lines are sensitive to probe the Milky Way's outer CGM ($r\gtrsim15$ kpc) and what ions are good tracers. We generate 5000 random QSO-star pair sightlines at $|b|>20\degree$, and compare the column density offsets among them for a list of ions, including \HI, \SiII, \CII, \SiIII, \SiIV, \CIV, \NV, \OVI, \NeVII, \NeVIII, \OVII, and \OVIII. The maximum angular separation allowed for the QSO-star pairs is set as 10$\degree$ with consideration of the simulation's resolution. We find that \OVI\ and \NV\ are likely to be good UV tracers of the Milky Way's outer CGM for their detectable $\delta \log N$ offsets between the QSOs and stars. \CIV\ may be less sensitive because the $\delta{\rm log}N$ excess is likely to be overwhelmed by the column density variations from sightline to sightline. 
    
\textbf{Bias IV} (\S\ref{sec:bias_in_out}): we show that the comparison between the Milky Way's CGM and extragalactic systems could be biased if the effect of different viewing angles is not taken into account. The Milky Way's CGM is observed from the inside, whereas extragalactic systems are observed externally. We find that the $\log N$ scatter from the inside-out views is a factor of $\sim 2$ higher than that from external views. 
The higher values from the inside-out views are because the lines of sight always have to pass through dense and clumpy gas in the inner halo. And, because bulk motions in the CGM occur mainly radially (Figures \ref{fig:phase_outflow}, \ref{fig:phase_inflow}), observations with inside-out views encounter more variances from sightline to sightline depending on whether or not the lines of sight intercept high-density inflows or low-density outflows.




\acknowledgments

We thank the anonymous referee for their constructive comments in improving our manuscript. We thank M.\ Putman, D.\ Weisz, A.\ Fox, N.\ Lehner, and J.\ C.\ Howk for their helpful feedback. Y.Z. acknowledges support from the Miller Institute for Basic Research in Science at the University of California, Berkeley. B.\ W.\ O.\
was supported in part by NSF grants PHY-1430152, AST-1514700, and OAC-1835213, by NASA grants NNX12AC98G
and NNX15AP39G, and by HST AR \#14315. 
R.\ A. and C.\ L.\ were supported by NASA grants 80NSSC18K1105, HST GO \# 14268, and HST GO \# 15075. RCS appreciates support from a Giacconi Fellowship at the Space Telescope Science Institute, which is operated by the Association of Universities for
Research in Astronomy, Inc., under NASA contract NAS 5-26555.
Resources supporting this work were provided by the NASA High-End Computing (HEC) Program through the NASA Advanced Supercomputing (NAS) Division at Ames Research Center and were sponsored by NASA's Science Mission Directorate; we are grateful for the superb user-support provided by NAS. Resources were also provided by the Blue Waters sustained-petascale computing project, which is supported by the NSF (award number ACI-1238993 and ACI-1514580) and the state of Illinois. Blue Waters is a joint effort of the University of Illinois at Urbana-Champaign and its NCSA.  This work benefited from the milky-way and haha emojis on Slack.   We have made extensive use of the python libraries Astropy, a community-developed core Python package for Astronomy (\citealp{astropy2}, \url{http://www.astropy.org}) and seaborn. Computations described in this work were performed using the publicly-available Enzo code, which is the product of a collaborative effort of many independent scientists from numerous institutions around the world.

\vspace{5mm}
\facilities{NASA Pleiades, NCSA Blue Waters}

\software{astropy \citep{astropy2},  
          Cloudy \citep{ferland13}, 
          Enzo \citep{bryan14,brummel19},
          grackle \citep{smith17},
          yt \citep{turk11},
          Trident \citep*{hummels17}
          }

\bibliographystyle{aasjournal}
\bibliography{references}

\end{document}